\magnification=\magstephalf
\newbox\SlashedBox
\def\slashed#1{\setbox\SlashedBox=\hbox{#1}
\hbox to 0pt{\hbox to 1\wd\SlashedBox{\hfil/\hfil}\hss}{#1}}
\def\hboxtosizeof#1#2{\setbox\SlashedBox=\hbox{#1}
\hbox to 1\wd\SlashedBox{#2}}

\def\mathslashed#1{\setbox\SlashedBox=\hbox{$#1$}
\hbox to 0pt{\hbox to 1\wd\SlashedBox{\hfil/\hfil}\hss}#1}

\def\ifsmall{\iffalse}  
\def\titlepagefont{}  

\def\DefineTeXgraphics{%
\special{ps::[global] /TeXgraphics { } def}}  

\def\today{\ifcase\month\or January\or February\or March\or April\or May
\or June\or July\or August\or September\or October\or November\or
December\fi\space\number\day, \number\year}
\def\eatPrefix19{}
\def\Year{\expandafter\eatPrefix\the\year}
\newcount\hours \newcount\minutes
\def\monthname{\ifcase\month\or
January\or February\or March\or April\or May\or June\or July\or
August\or September\or October\or November\or December\fi}
\def\shortmonthname{\ifcase\month\or
Jan\or Feb\or Mar\or Apr\or May\or Jun\or Jul\or
Aug\or Sep\or Oct\or Nov\or Dec\fi}

\def\TimeStamp{\hours\the\time\divide\hours by60%
\minutes -\the\time\divide\minutes by60\multiply\minutes by60%
\advance\minutes by\the\time%
${\rm \shortmonthname}\cdot\if\day<10{}0\fi\the\day\cdot\the\year%
\qquad\the\hours:\if\minutes<10{}0\fi\the\minutes$}




\def\Title#1{%
\vskip 1in{\titlefont\centerline{#1}}\vskip .5in}



\newif\ifdraftmode
\newif\ifleftlabels  

\def\nolabels{\def\wrlabeL##1{}\def\eqlabeL##1{}\def\reflabeL##1{}}
\def\writelabels{\def\wrlabeL##1{\leavevmode\vadjust{\rlap{\smash%
{\line{{\escapechar=` \hfill\rlap{\sevenrm\hskip.03in\string##1}}}}}}}%
\def\eqlabeL##1{{\escapechar-1\rlap{\sevenrm\hskip.05in\string##1}}}%
\def\reflabeL##1{\noexpand\rlap{\noexpand\sevenrm[\string##1]}}}
\def\writeleftlabels{\def\wrlabeL##1{\leavevmode\vadjust{\rlap{\smash%
{\line{{\escapechar=` \hfill\rlap{\sevenrm\hskip.03in\string##1}}}}}}}%
\def\eqlabeL##1{{\escapechar-1%
\rlap{\sixrm\hskip.05in\string##1}%
\llap{\sevenrm\string##1\hskip.03in\hbox to \hsize{}}}}%
\def\reflabeL##1{\noexpand\rlap{\noexpand\sevenrm[\string##1]}}}
\nolabels

\newdimen\fullhsize
\newdimen\hstitle
\hstitle=\hsize 
\newdimen\hsbody
\hsbody=\hsize 
\newdimen\hbodyoffset
\hbodyoffset=\hoffset 
\newbox\leftpage
\def\abstract#1{#1}
\def\rotated{\special{ps: landscape}
\magnification=1000  
\baselineskip=14pt
\global\hstitle=9truein\global\hsbody=4.75truein
\global\vsize=7truein\global\voffset=-.31truein
\global\hoffset=-0.54in\global\hbodyoffset=-.54truein
\global\fullhsize=10truein
\def\DefineTeXgraphics{%
\special{ps::[global]
/TeXgraphics {currentpoint translate 0.7 0.7 scale
              -80 0.72 mul -1000 0.72 mul translate} def}}
\let\lr=L
\def\ifsmall{\iftrue}
\def\titlepagefont{\twelvepoint}
\trueseventeenpoint
\def\almostshipout##1{\if L\lr \count1=1
      \global\setbox\leftpage=##1 \global\let\lr=R
   \else \count1=2
      \shipout\vbox{\hbox to\fullhsize{\box\leftpage\hfil##1}}
      \global\let\lr=L\fi}

\output={\ifnum\count0=1 
 \shipout\vbox{\hbox to \fullhsize{\hfill\pagebody\hfill}}\advancepageno
 \else
 \almostshipout{\leftline{\vbox{\pagebody\makefootline}}}\advancepageno
 \fi}

\def\abstract##1{{\leftskip=1.5in\rightskip=1.5in ##1\par}} }

\def\linemessage#1{\immediate\write16{#1}}

\global\newcount\secno \global\secno=0
\global\newcount\appno \global\appno=0
\global\newcount\meqno \global\meqno=1
\global\newcount\subsecno \global\subsecno=0
\global\newcount\figno \global\figno=0

\newif\ifAnyCounterChanged
\let\terminator=\relax
\def\normalize#1{\ifx#1\terminator\let\next=\relax\else%
\if#1i\aftergroup i\else\if#1v\aftergroup v\else\if#1x\aftergroup x%
\else\if#1l\aftergroup l\else\if#1c\aftergroup c\else%
\if#1m\aftergroup m\else%
\if#1I\aftergroup I\else\if#1V\aftergroup V\else\if#1X\aftergroup X%
\else\if#1L\aftergroup L\else\if#1C\aftergroup C\else%
\if#1M\aftergroup M\else\aftergroup#1\fi\fi\fi\fi\fi\fi\fi\fi\fi\fi\fi\fi%
\let\next=\normalize\fi%
\next}
\def\makeNormal#1#2{\def\doNormalDef{\edef#1}\begingroup%
\aftergroup\doNormalDef\aftergroup{\normalize#2\terminator\aftergroup}%
\endgroup}

\def\warnIfChanged#1#2{%
\ifundef#1
\else\begingroup%
\edef\oldDefinitionOfCounter{#1}\edef\newDefinitionOfCounter{#2}%
\ifx\oldDefinitionOfCounter\newDefinitionOfCounter%
\else%
\linemessage{Warning: definition of \noexpand#1 has changed.}%
\global\AnyCounterChangedtrue\fi\endgroup\fi}

\def\Section#1{\global\advance\secno by1\relax\global\meqno=1%
\global\subsecno=0%
\bigbreak\bigskip
\centerline{\twelvepoint \bf %
\the\secno. #1}%
\par\nobreak\medskip\nobreak}
\def\tagsection#1{%
\warnIfChanged#1{\the\secno}%
\xdef#1{\the\secno}%
\ifWritingAuxFile\immediate\write\auxfile{\noexpand\xdef\noexpand#1{#1}}\fi%
}
\def\section{\Section}
\def\Subsection#1{\global\advance\subsecno by1\relax\medskip %
\leftline{\bf\the\secno.\the\subsecno\ #1}%
\par\nobreak\smallskip\nobreak}
\def\tagsubsection#1{%
\warnIfChanged#1{\the\secno.\the\subsecno}%
\xdef#1{\the\secno.\the\subsecno}%
\ifWritingAuxFile\immediate\write\auxfile{\noexpand\xdef\noexpand#1{#1}}\fi%
}

\def\subsection{\Subsection}

\def\romappno{\uppercase\expandafter{\romannumeral\appno}}
\def\makeNormalizedRomappno{%
\expandafter\makeNormal\expandafter\normalizedromappno%
\expandafter{\romannumeral\appno}%
\edef\normalizedromappno{\uppercase{\normalizedromappno}}}
\def\Appendix#1{\global\advance\appno by1\relax\global\meqno=1\global\secno=0%
\global\subsecno=0%
\bigbreak\bigskip
\centerline{\twelvepoint \bf Appendix %
\romappno. #1}%
\par\nobreak\medskip\nobreak}
\def\tagappendix#1{\makeNormalizedRomappno%
\warnIfChanged#1{\normalizedromappno}%
\xdef#1{\normalizedromappno}%
\ifWritingAuxFile\immediate\write\auxfile{\noexpand\xdef\noexpand#1{#1}}\fi%
}
\def\appendix{\Appendix}
\def\Subappendix#1{\global\advance\subsecno by1\relax\medskip %
\leftline{\bf\romappno.\the\subsecno\ #1}%
\par\nobreak\smallskip\nobreak}
\def\tagsubappendix#1{\makeNormalizedRomappno%
\warnIfChanged#1{\normalizedromappno.\the\subsecno}%
\xdef#1{\normalizedromappno.\the\subsecno}%
\ifWritingAuxFile\immediate\write\auxfile{\noexpand\xdef\noexpand#1{#1}}\fi%
}

\def\eqn#1{\makeNormalizedRomappno%
\ifnum\secno>0%
  \warnIfChanged#1{\the\secno.\the\meqno}%
  \eqno(\the\secno.\the\meqno)\xdef#1{\the\secno.\the\meqno}%
     \global\advance\meqno by1
\else\ifnum\appno>0%
  \warnIfChanged#1{\normalizedromappno.\the\meqno}%
  \eqno({\rm\romappno}.\the\meqno)%
      \xdef#1{\normalizedromappno.\the\meqno}%
     \global\advance\meqno by1
\else%
  \warnIfChanged#1{\the\meqno}%
  \eqno(\the\meqno)\xdef#1{\the\meqno}%
     \global\advance\meqno by1
\fi\fi%
\eqlabeL#1%
\ifWritingAuxFile\immediate\write\auxfile{\noexpand\xdef\noexpand#1{#1}}\fi%
}
\def\defeqn#1{\makeNormalizedRomappno%
\ifnum\secno>0%
  \warnIfChanged#1{\the\secno.\the\meqno}%
  \xdef#1{\the\secno.\the\meqno}%
     \global\advance\meqno by1
\else\ifnum\appno>0%
  \warnIfChanged#1{\normalizedromappno.\the\meqno}%
  \xdef#1{\normalizedromappno.\the\meqno}%
     \global\advance\meqno by1
\else%
  \warnIfChanged#1{\the\meqno}%
  \xdef#1{\the\meqno}%
     \global\advance\meqno by1
\fi\fi%
\eqlabeL#1%
\ifWritingAuxFile\immediate\write\auxfile{\noexpand\xdef\noexpand#1{#1}}\fi%
}
\def\anoneqn{\makeNormalizedRomappno%
\ifnum\secno>0
  \eqno(\the\secno.\the\meqno)%
     \global\advance\meqno by1
\else\ifnum\appno>0
  \eqno({\rm\normalizedromappno}.\the\meqno)%
     \global\advance\meqno by1
\else
  \eqno(\the\meqno)%
     \global\advance\meqno by1
\fi\fi%
}
\def\mfig#1#2{\ifx#20
\else\global\advance\figno by1%
\relax#1\the\figno%
\warnIfChanged#2{\the\figno}%
\xdef#2{\the\figno}%
\reflabeL#2%
\ifWritingAuxFile\immediate\write\auxfile{\noexpand\xdef\noexpand#2{#2}}\fi\fi%
}

\catcode`@=11 

\newif\ifFiguresInText\FiguresInTexttrue
\newif\if@FigureFileCreated
\newwrite\capfile
\newwrite\figfile

\newif\ifcaption
\captiontrue
\def\captionsize{\tenrm}
\def\PlaceTextFigure#1#2#3#4{%
\vskip 0.5truein%
#3\hfil\epsfbox{#4}\hfil\break%
\ifcaption\hfil\vbox{\captionsize Figure #1. #2}\hfil\fi%
\vskip10pt}
\def\PlaceEndFigure#1#2{%
\epsfxsize=\hsize\epsfbox{#2}\vfill\centerline{Figure #1.}\eject}

\def\LoadFigure#1#2#3#4{%
\ifundef#1{\phantom{\mfig{}#1}}\else
\fi%
\ifFiguresInText
\PlaceTextFigure{#1}{#2}{#3}{#4}%
\else
\if@FigureFileCreated\else%
\immediate\openout\capfile=\jobname.caps%
\immediate\openout\figfile=\jobname.figs%
@FigureFileCreatedtrue\fi%
\immediate\write\capfile{\noexpand\item{Figure \noexpand#1.\ }{#2}\vskip10pt}%
\immediate\write\figfile{\noexpand\PlaceEndFigure\noexpand#1{\noexpand#4}}%
\fi}

\def\listfigs{\ifFiguresInText\else%
\vfill\eject\immediate\closeout\capfile
\immediate\closeout\figfile%
\centerline{{\bf Figures}}\bigskip\frenchspacing%
\catcode`@=11 
\def\captionsize{\tenrm}
\input \jobname.caps\vfill\eject\nonfrenchspacing%
\catcode`\@=\active
\catcode`@=12  
\input\jobname.figs\fi}

\font\ninerm=cmr9
\font\eightrm=cmr8
\font\sixrm=cmr6

\def\loadtrueseventeenpoint{
 \font\seventeenrm=cmr10 at 17.28truept
 \font\seventeeni=cmmi10 at 17.28truept
 \font\seventeenbf=cmbx10 at 17.28truept
 \font\seventeenit=cmti10 at 17.28truept
 \font\seventeensl=cmsl10 at 17.28truept
 \font\seventeensy=cmsy10 at 17.28truept
}
\def\loadfourteenpoint{
\font\fourteenrm=cmr10 at 14.4pt
\font\fourteeni=cmmi10 at 14.4pt
\font\fourteenit=cmti10 at 14.4pt
\font\fourteensl=cmsl10 at 14.4pt
\font\fourteensy=cmsy10 at 14.4pt
\font\fourteenbf=cmbx10 at 14.4pt
}
\def\loadtruetwelvepoint{
\font\twelverm=cmr10 at 12truept
\font\twelvei=cmmi10 at 12truept
\font\twelveit=cmti10 at 12truept
\font\twelvesl=cmsl10 at 12truept
\font\twelvesy=cmsy10 at 12truept
\font\twelvebf=cmbx10 at 12truept
}

\font\ninei=cmmi9
\font\eighti=cmmi8
\font\sixi=cmmi6
\skewchar\ninei='177 \skewchar\eighti='177 \skewchar\sixi='177

\font\ninesy=cmsy9
\font\eightsy=cmsy8
\font\sixsy=cmsy6
\skewchar\ninesy='60 \skewchar\eightsy='60 \skewchar\sixsy='60

\font\ninebf=cmbx9
\font\eightbf=cmbx8
\font\sixbf=cmbx6

\font\ninett=cmtt9
\font\eighttt=cmtt8

\hyphenchar\tentt=-1 
\hyphenchar\ninett=-1
\hyphenchar\eighttt=-1

\font\ninesl=cmsl9
\font\eightsl=cmsl8

\font\nineit=cmti9
\font\eightit=cmti8


\newskip\ttglue
\def\tenpoint{\def\rm{\fam0\tenrm}%
  \textfont0=\tenrm \scriptfont0=\sevenrm \scriptscriptfont0=\fiverm
  \textfont1=\teni \scriptfont1=\seveni \scriptscriptfont1=\fivei
  \textfont2=\tensy \scriptfont2=\sevensy \scriptscriptfont2=\fivesy
  \textfont3=\tenex \scriptfont3=\tenex \scriptscriptfont3=\tenex
  \def\it{\fam\itfam\tenit}\textfont\itfam=\tenit
  \def\sl{\fam\slfam\tensl}\textfont\slfam=\tensl
  \def\bf{\fam\bffam\tenbf}\textfont\bffam=\tenbf \scriptfont\bffam=\sevenbf
  \scriptscriptfont\bffam=\fivebf
  \normalbaselineskip=12pt
  \let\sc=\eightrm
  \let\big=\tenbig
  \setbox\strutbox=\hbox{\vrule height8.5pt depth3.5pt width\z@}%
  \normalbaselines\rm}

\def\twelvepoint{\def\rm{\fam0\twelverm}%
  \textfont0=\twelverm \scriptfont0=\ninerm \scriptscriptfont0=\sevenrm
  \textfont1=\twelvei \scriptfont1=\ninei \scriptscriptfont1=\seveni
  \textfont2=\twelvesy \scriptfont2=\ninesy \scriptscriptfont2=\sevensy
  \textfont3=\tenex \scriptfont3=\tenex \scriptscriptfont3=\tenex
  \def\it{\fam\itfam\twelveit}\textfont\itfam=\twelveit
  \def\sl{\fam\slfam\twelvesl}\textfont\slfam=\twelvesl
  \def\bf{\fam\bffam\twelvebf}\textfont\bffam=\twelvebf%
  \scriptfont\bffam=\ninebf
  \scriptscriptfont\bffam=\sevenbf
  \normalbaselineskip=12pt
  \let\sc=\eightrm
  \let\big=\tenbig
  \setbox\strutbox=\hbox{\vrule height8.5pt depth3.5pt width\z@}%
  \normalbaselines\rm}

\def\fourteenpoint{\def\rm{\fam0\fourteenrm}%
  \textfont0=\fourteenrm \scriptfont0=\tenrm \scriptscriptfont0=\sevenrm
  \textfont1=\fourteeni \scriptfont1=\teni \scriptscriptfont1=\seveni
  \textfont2=\fourteensy \scriptfont2=\tensy \scriptscriptfont2=\sevensy
  \textfont3=\tenex \scriptfont3=\tenex \scriptscriptfont3=\tenex
  \def\it{\fam\itfam\fourteenit}\textfont\itfam=\fourteenit
  \def\sl{\fam\slfam\fourteensl}\textfont\slfam=\fourteensl
  \def\bf{\fam\bffam\fourteenbf}\textfont\bffam=\fourteenbf%
  \scriptfont\bffam=\tenbf
  \scriptscriptfont\bffam=\sevenbf
  \normalbaselineskip=17pt
  \let\sc=\elevenrm
  \let\big=\tenbig
  \setbox\strutbox=\hbox{\vrule height8.5pt depth3.5pt width\z@}%
  \normalbaselines\rm}

\def\seventeenpoint{\def\rm{\fam0\seventeenrm}%
  \textfont0=\seventeenrm \scriptfont0=\fourteenrm \scriptscriptfont0=\tenrm
  \textfont1=\seventeeni \scriptfont1=\fourteeni \scriptscriptfont1=\teni
  \textfont2=\seventeensy \scriptfont2=\fourteensy \scriptscriptfont2=\tensy
  \textfont3=\tenex \scriptfont3=\tenex \scriptscriptfont3=\tenex
  \def\it{\fam\itfam\seventeenit}\textfont\itfam=\seventeenit
  \def\sl{\fam\slfam\seventeensl}\textfont\slfam=\seventeensl
  \def\bf{\fam\bffam\seventeenbf}\textfont\bffam=\seventeenbf%
  \scriptfont\bffam=\fourteenbf
  \scriptscriptfont\bffam=\twelvebf
  \normalbaselineskip=21pt
  \let\sc=\fourteenrm
  \let\big=\tenbig
  \setbox\strutbox=\hbox{\vrule height 12pt depth 6pt width\z@}%
  \normalbaselines\rm}

\def\ninepoint{\def\rm{\fam0\ninerm}%
  \textfont0=\ninerm \scriptfont0=\sixrm \scriptscriptfont0=\fiverm
  \textfont1=\ninei \scriptfont1=\sixi \scriptscriptfont1=\fivei
  \textfont2=\ninesy \scriptfont2=\sixsy \scriptscriptfont2=\fivesy
  \textfont3=\tenex \scriptfont3=\tenex \scriptscriptfont3=\tenex
  \def\it{\fam\itfam\nineit}\textfont\itfam=\nineit
  \def\sl{\fam\slfam\ninesl}\textfont\slfam=\ninesl
  \def\bf{\fam\bffam\ninebf}\textfont\bffam=\ninebf \scriptfont\bffam=\sixbf
  \scriptscriptfont\bffam=\fivebf
  \normalbaselineskip=11pt
  \let\sc=\sevenrm
  \let\big=\ninebig
  \setbox\strutbox=\hbox{\vrule height8pt depth3pt width\z@}%
  \normalbaselines\rm}

\def\eightpoint{\def\rm{\fam0\eightrm}%
  \textfont0=\eightrm \scriptfont0=\sixrm \scriptscriptfont0=\fiverm%
  \textfont1=\eighti \scriptfont1=\sixi \scriptscriptfont1=\fivei%
  \textfont2=\eightsy \scriptfont2=\sixsy \scriptscriptfont2=\fivesy%
  \textfont3=\tenex \scriptfont3=\tenex \scriptscriptfont3=\tenex%
  \def\it{\fam\itfam\eightit}\textfont\itfam=\eightit%
  \def\sl{\fam\slfam\eightsl}\textfont\slfam=\eightsl%
  \def\bf{\fam\bffam\eightbf}\textfont\bffam=\eightbf \scriptfont\bffam=\sixbf%
  \scriptscriptfont\bffam=\fivebf%
  \normalbaselineskip=9pt%
  \let\sc=\sixrm%
  \let\big=\eightbig%
  \setbox\strutbox=\hbox{\vrule height7pt depth2pt width\z@}%
  \normalbaselines\rm}

\def\tenbig#1{{\hbox{$\left#1\vbox to8.5pt{}\right.\n@space$}}}
\def\ninebig#1{{\hbox{$\textfont0=\tenrm\textfont2=\tensy
  \left#1\vbox to7.25pt{}\right.\n@space$}}}
\def\eightbig#1{{\hbox{$\textfont0=\ninerm\textfont2=\ninesy
  \left#1\vbox to6.5pt{}\right.\n@space$}}}

\def\footnote#1{\edef\@sf{\spacefactor\the\spacefactor}#1\@sf
      \insert\footins\bgroup\eightpoint
      \interlinepenalty100 \let\par=\endgraf
        \leftskip=\z@skip \rightskip=\z@skip
        \splittopskip=10pt plus 1pt minus 1pt \floatingpenalty=20000
        \smallskip\item{#1}\bgroup\strut\aftergroup\@foot\let\next}
\skip\footins=12pt plus 2pt minus 4pt 
\dimen\footins=30pc 

\newinsert\margin
\dimen\margin=\maxdimen
\def\titlefont{\seventeenpoint}
\loadtruetwelvepoint 
\loadtrueseventeenpoint

\def\eatOne#1{}
\def\ifundef#1{\expandafter\ifx%
\csname\expandafter\eatOne\string#1\endcsname\relax}
\def\notTrue{\iffalse}\def\isTrue{\iftrue}
\def\ifdef#1{{\ifundef#1%
\aftergroup\notTrue\else\aftergroup\isTrue\fi}}
\def\use#1{\ifundef#1\linemessage{Warning: \string#1 is undefined.}%
{\tt \string#1}\else#1\fi}



%
\catcode`"=11
\let\quote="
\catcode`"=12
\chardef\foo="22
\global\newcount\refno \global\refno=1
\newwrite\rfile
\newlinechar=`\^^J
\def\@ref#1#2{\the\refno\n@ref#1{#2}}
\def\h@ref#1#2#3{\href{#3}{\the\refno}\n@ref#1{#2}}
\def\n@ref#1#2{\xdef#1{\the\refno}%
\ifnum\refno=1\immediate\openout\rfile=\jobname.refs\fi%
\immediate\write\rfile{\noexpand\item{[\noexpand#1]\ }#2.}%
\global\advance\refno by1}
\def\nref{\n@ref} 
\def\ref{\@ref}   
\def\hrref{\h@ref}
\def\lref#1#2{\the\refno\xdef#1{\the\refno}%
\ifnum\refno=1\immediate\openout\rfile=\jobname.refs\fi%
\immediate\write\rfile{\noexpand\item{[\noexpand#1]\ }#2\semi}%
\global\advance\refno by1}
\def\cref#1{\immediate\write\rfile{#1\semi}}

\def\preref#1#2{\gdef#1{\@ref#1{#2}}}

\def\semi{;\hfil\noexpand\break}

\def\listrefs{\vfill\eject\immediate\closeout\rfile
\centerline{{\bf References}}\bigskip\frenchspacing%
\input \jobname.refs\vfill\eject\nonfrenchspacing}

\def\inputAuxIfPresent#1{\immediate\openin1=#1
\ifeof1\message{No file \auxfileName; I'll create one.
}\else\closein1\relax\input\auxfileName\fi%
}




\newif\ifWritingAuxFile
\newwrite\auxfile
\def\SetUpAuxFile{%
\xdef\auxfileName{\jobname.aux}%
\inputAuxIfPresent{\auxfileName}%
\WritingAuxFiletrue%
\immediate\openout\auxfile=\auxfileName}

\def\L{\left(}\def\R{\right)}
\def\LP{\left.}\def\RP{\right.}
\def\LB{\left[}\def\RB{\right]}

\def\RV{\right|}

\def\bye{\par\vfill\supereject%
\ifAnyCounterChanged\linemessage{
Some counters have changed.  Re-run tex to fix them up.}\fi%
\end}

\catcode`\@=\active
\catcode`@=12  
\catcode`\"=\active


\hfuzz 30 pt
\overfullrule 0pt
\SetUpAuxFile

 \loadfourteenpoint
\font\smc=cmcsc10

\noindent

\def\Re{\mathop{\rm Re}}
\def\Im{\mathop{\rm Im}}
\def\Tr{\mathop{\rm Tr}}
\def\F#1#2{\,{{\vphantom{F}}_{#1}F_{#2}}}

\preref\DR{P. J. Davis and P. Rabinowitz,
 {\it Methods of Numerical Integration\/}, 2nd Edition, Academic Press, 1984}
\preref\GR{M. Gl{\accent 127 u}ck and E. Reya, Phys.\ Rev.\ D14:3034 (1976)}
\preref\GRV{M. Gl{\accent 127 u}ck, E. Reya, and A. Vogt, Z. Phys.\ C48:471
(1990)}
\preref\MRS{A. D. Martin, R. G. Roberts, and W. J. Stirling,
Phys.\ Lett.\ B387:419 (1996) [hep-ph/9606345];
Phys.\ Lett.\ B354:155 (1995) [hep-ph/9502336];
Int.\ J.\ Mod.\ Phys.\ A10:2885 (1995)}
\preref\CTEQ{H. L. Lai, J. Huston, S. Kuhlmann, F. Olness, J. F. Owens, D.
Soper,
W. K. Tung, H. Weerts, Phys.\ Rev.\ D55:1280 (1997) [hep-ph/9606399]\semi
H. L. Lai, J. Botts, J. Huston, J. G. Morfin, J. F. Owens, J. W. Qiu, W. K.
Tung,
H. Weerts, Phys.\ Rev.\ D51:4763 (1995) [hep-ph/9410404]\semi
G. Sterman, J. Smith, J. C. Collins, J. Whitmore, R. Brock, J. Huston,
J. Pumplin, W.-K. Tung, H. Weerts, C.-P. Yuan, S. Kuhlmann,
S. Mishra, J. G. Morfin, F. Olness, J. Owens, J. Qiu, and D. E. Soper,
Rev.\ Mod.\ Phys.\ 67:157 (1995)}
\preref\EvolEqnRef{G. Altarelli and G. Parisi, Nucl.\ Phys.\ B126:298 (1977)}
\preref\Dokshitser{Yu. L. Dokshitser, Sov.\ Phys.\ JETP 46:641 (1977)}
\preref\BetaRefs{D. J. Gross and F. W. Wilczek, Phys.\ Rev.\ Lett.\ 30:1343
(1973)\semi
                 H. D. Politzer, Phys.\ Rev.\ Lett.\ 30:1346 (1973)\semi
                 W. Caswell, Phys.\ Rev.\ Lett.\ 33:224 (1974)\semi
                 D. R. T. Jones, Nucl.\ Phys.\ B87:127 (1975)}
\preref\Blum{J. Bl{\accent 127 u}mlein, S. Riemersma, W. L. van Neerven, and A.
Vogt,
             Nucl.\ Phys.\ Proc.\ Suppl.\ 51C:97 (1996)
             [hep-ph/9609217]}
\preref\Compare{J. Bl{\accent 127 u}mlein, S. Riemersma, M. Botje, C. Pascaud,
F. Zomer,
                W. L. van Neerven, A. Vogt, hep-ph/9609400}
\preref\Mellin{M. Diemoz, F. Ferroni, E. Longo, and G. Martinelli,
               Z. Phys.\ C39:21 (1988)}
\preref\Direct{M. Virchaux and A. Ouraou, DPhPE 87--15\semi
               M. Virchaux, PhD Thesis, University of Paris--7 (1988)\semi
               A. Ouraou, PhD Thesis, University of Paris--11 (1988)\semi
               M. Botje, {\smc qcdnum}15: {\it A fast QCD evolution program\/},
                 to appear\semi
               C. Pascaud and F. Zomer, HERA H1 note H1--11/94--404}
\preref\FP{W. Furmanski and R. Petronzio, Z. Phys.\ C11:293 (1982)}
\preref\FKL{E. G. Floratos, C. Kounnas, and R. Lacaze,
Nucl. Phys. B192:417 (1981)}
\preref\GGK{W. T. Giele, E. W. N. Glover, and D. A. Kosower, Nucl. Phys.
B403:633 (1993)
[hep-ph/9302225]}
\preref\Extraction{D. A. Kosower, to appear}
$\null$

\vskip -.6 cm

\nopagenumbers
\noindent hep-th/9706213 \hfill {Saclay--SPhT/T97--43}

\vskip -2.0 cm

\baselineskip 12 pt
\Title{\bf Evolution of Parton Distributions}

\vskip 1.0truein

\centerline{\ninerm David A. Kosower}
\baselineskip12truept
\centerline{\nineit Service de Physique Th\'eorique${}^{\dagger}$,
Centre d'Etudes de Saclay}
\centerline{\nineit F--91191 Gif-sur-Yvette cedex, France}
\centerline{\tt kosower@spht.saclay.cea.fr}

\vskip 0.5truein
\vglue  0.3cm

\vskip 0.2truein
\baselineskip13truept
\centerline{\bf Abstract}

{\narrower

I present a highly efficient method for evolving parton distributions
in perturbative QCD.  The method allows evolving the parton distribution
functions according to any of the commonly-used truncations
of the evolution equations (which differ in their treatment
of higher-order terms).  I also give formul\ae\ for computing crossing
functions within the method.
}

\vskip 3.7 cm
\centerline{\it Submitted to Nuclear Physics B}

\vfil\vskip .2 cm
\noindent\hrule width 3.6in\hfil\break
${}^{\dagger}$Laboratory of the {\it Direction des Sciences de
la Mati\`ere\/}
of the {\it Commissariat \`a l'Energie Atomique\/} of
France.\hfil\break
       \eject
\footline={\hss\tenrm\folio\hss}

\baselineskip17pt

\noindent
\section{Introduction}
\vskip 10pt

The parton distribution functions of the nucleon are
fundamental ingredients in the description of short-distance
scattering experiments involving hadronic initial states,
whether these be in hadron-hadron collisions,
or in deeply inelastic scattering.  Indeed, along with
the running coupling $\alpha_s$, they are the only ingredients
needed from outside perturbation theory for a complete description of
such processes ignoring subleading power corrections.

Both the parton distributions and the running coupling depend
on the momentum scale at which they are evaluated.  As is well known, the
{\it variation\/} in each of these quantities as one moves
from one momentum scale to another is described by an evolution
equation that can be computed perturbatively.  It is only the
values of the distributions and of the coupling at a fixed scale
which are required inputs from outside perturbation theory.

Lattice gauge theorists may calculate these quantities someday, but
for the moment they must be extracted from experiments.  The
modern approaches~[\use\MRS,\use\CTEQ] involve global fits
to all available experiments.  The experiments in fact involve different
scale arguments to the distribution functions, but as these are related
by a perturbative evolution equation, we can regard the fits as determining
the distributions at a certain fixed scale $Q_0$.

Present theoretical abilities allow extensive calculations at
next-to-leading order (NLO), and there are growing indications that
next-to-next-to-leading order (NNLO) calculations are required if
we are to determine $\alpha_s$ to 1\%.  To the required
precision, the evolution
equation for the running coupling can be solved in closed form.  This is not
true, however, of the evolution equation for the parton distributions;
these must be evolved numerically from the original fixed scale.

Workers have used two main approaches to evolve the distribution
functions numerically: direct integration of the evolution equations,
and an approach based on Mellin transformations.  This paper presents
an improvement to the techniques heretofore used, within the framework
of the latter approach.

\section{Evolution Equations}
\tagsection\EvolutionSection
\vskip 10pt

A parton distribution function $f(x,Q^2)$ obeys an evolution
equation~[\use\EvolEqnRef]
in momentum,
$$\eqalign{
Q^2 {\partial f(x,Q^2)\over \partial Q^2} = P(x,Q^2) \otimes f(x,Q^2)\,,
}\eqn\OriginalEvolution$$
where $P(x,Q^2)$ is the Altarelli-Parisi kernel,
and $\otimes$ denotes a convolution,
$$
A(x) \otimes B(x) = \int_0^1 dy \int_0^1 dz\;
\,\delta(x-yz)A(y) B(z) \,,
\anoneqn$$
(For a different early discussion, see ref.~[\use\Dokshitser].)
For the quark non-singlet distributions, each of the quantities in
eqn.~(\use\OriginalEvolution) is a scalar; in the singlet sector,
$f$ is a two-component vector containing the quark singlet distribution
and the gluon distribution, and $P(x,Q^2)$ is a $2\times2$ matrix.

The kernel $P(x,Q^2)$ has a perturbative expansion,
$$
P(x,Q^2) = a_s(Q^2) P_0(x) + a_s^2(Q^2) P_1(x) + {\cal O}(a_s^3)\,,
\anoneqn$$
in which $a_s(Q^2) = \alpha_s(Q^2)/(4\pi)$
is a rescaled version of the usual running coupling. As mentioned
in the introduction, one approach to evolving the parton distribution
functions $f$, known as the $x$-space method,
 is to integrate the perturbative approximation
to eqn.~(\use\OriginalEvolution) directly.  I will
return to the issues surrounding the choice of method in
section~\use\XSpaceSection.  For the moment, let us proceed by Mellin
transforming the evolution equation; with Mellin moments $h^z$ of a
function $h(x)$ defined via
$$
h^z = \int_0^1 dx\; x^{z-1} h(x)\,,
\anoneqn$$
the evolution equation, truncated to next-to-leading order, becomes
$$
Q^2 {\partial f^z(Q^2)\over\partial Q^2} =
  \LB a_s(Q^2) P_0^z + a_s^2(Q^2) P_1^z \RB\, f^z(Q^2)
\eqn\MellinFullEquation$$
since the Mellin transformation turns convolutions into products.  (That is
reason for using it.)  It is convenient to change variables, and to use
$a_s$ as the evolution variable.  We can do this using the beta function,
$$
\beta(a_s) \equiv {\partial a_s\over \partial \ln Q^2}
  =-\beta_0 a_s^2 -\beta_1 a_s^3 + {\cal O}(a_s^4)\,,
\eqn\BetaFunction$$
obtaining
$$
{\partial f^z\over\partial a_s}
 = -{\LB P_0^z + a_s P_1^z \RB\over
    a_s\LB\beta_0+\beta_1 a_s\RB }\, f^z\,.
\eqn\AsEqn$$
The usual approach~[\use\Blum] expands the right-hand side, in keeping with
a perturbative treatment of the equation, upon which we obtain
$$
{\partial f^z\over\partial a_s}
 = -{1\over\beta_0 a_s}\,
    \LB P_0^z + a_s \L P_1^z-{\beta_1\over\beta_0} P_0^z\R \RB\, f^z\,.
\eqn\TruncatedEqn$$

The boundary condition is $f^z(Q_0^2) = f_0^z$.  Conventionally one
proceeds by introducing an evolution operator $E$ such that
$$
f^z(Q^2) = E^z(a_s(Q^2),a_s(Q_0^2)) f^z(Q_0^2)\,.
\eqn\EvolutionOperatorDef$$
In the non-singlet case, $E$ is just a scalar function of $z$;
in the singlet case, it is a $2\times2$ matrix.  The evolution operator
satisfies the same equation~(\use\TruncatedEqn) we have been considering above,
$$
{\partial E^z\over\partial a_s}
 = -{1\over\beta_0 a_s}\,
    \LB P_0^z + a_s \L P_1^z-{\beta_1\over\beta_0} P_0^z\R \RB\, E^z\,,
\eqn\TruncatedEvolution$$
but has a boundary condition, $E^z(a_0,a_0) = 1$, that renders it an
entirely perturbative object.

\def\qbar{{\overline q}}\def\ubar{{\overline u}}\def\dbar{{\overline d}}
\def\Pb{{\overline P}}
Solving the evolution equation is straightforward in the non-singlet
case; expanding in $a_s$ beyond the leading terms,
we obtain~[\use\FP,\use\GRV]
$$
E^z_\eta(a_s,a_0) =
 \LB 1- {a_s-a_0\over\beta_0} \Pb_1^z(\eta)
    \RB\,\L{a_s\over a_0}\R^{-P_0^z/\beta_0}\,.
\eqn\EvolutionOperator$$
where $\eta=\pm1$ correspond respectively to the combinations
$q-\qbar$ and $\{(u+\ubar)-(d+\dbar),\ldots\}$, and where
$\Pb_1^z(\eta)\equiv P_1^z(\eta) - {\beta_1\over\beta_0}P_0^z$.
In the singlet case, we need first to define two matrices which project
onto the eigenvectors of the leading-order AP coefficient matrix,
$$\eqalign{
P_{\pm}^z &= \pm{1\over\L \lambda^z_{+}-\lambda^z_{-}\R}
               \L P_0^z -\lambda^z_{\mp}\R\,,
\cr
}$$
in which
$$\eqalign{
\lambda^z_{\pm} &=
{1\over2}\LB P_{0,gg}^{z} + P_{0,qq}^{z}
  \pm \sqrt{\L P_{0,gg}^{z}-P_{0,qq}^{z}\R^2
                      +4 P_{0,qg}^{z} P_{0,gq}^{z}}\RB\,\cr
}\eqn\LambdaDefinitions$$
are the eigenvalues of $P_0^z$.  With these definitions, we can
write the evolution operator in the singlet sector as~[\use\FP,\use\GRV],
$$\eqalign{
E^z(a_s,a_0) &=
 \LB P^z_{-}
 -{a_s-a_0\over\beta_0} P_{-}^z \Pb_1^z P_{-}^z
\vphantom{ \L a_s\L{a_s\over a_0}\R^{(\lambda_{-}^z)/\beta_0}-a_0\R } \RP\cr
&\hskip 5mm \LP
- {1\over \beta_0-\lambda_{+}^z+\lambda_{-}^z}
\L a_s \L{a_s\over a_0}\R^{(\lambda_{-}^z-\lambda_{+}^z)/\beta_0}-a_0 \R\,
     P_{-}^z \Pb_1^z P_{+}^z
    \RB\,\L{a_s\over a_0}\R^{-\lambda_{-}^z/\beta_0}\cr
&\hskip 15mm
\vphantom{ \L a_s\L{a_s\over a_0}\R^{(\lambda_{-}^z)/\beta_0}-a_0\R }
 + \L +\leftrightarrow -\R\,.
}\eqn\SingletEvolutionOperator$$

With the normalizations used here and $n_f$ the number of active flavors,
the beta function~[\use\BetaRefs] coefficients are
$$\beta_0 = 11 - {2\over3} n_f,\qquad \beta_1 = 102 - {38\over3} n_f\,;
\eqn\BetaFunctionCoefficients$$
the Mellin moments of the Altarelli-Parisi function may be found
in ref.~[\use\FKL],
noting that $P_i^z = -\gamma^{(i) z}/2$.  The analytic continuations of
these moments
are given by Gl\"uck, Reya, and Vogt [\use\GRV]; I discuss the analytic
continuation of a certain term involving the dilogarithm in
an appendix.

Of course, for use in perturbative cross section calculations, we want
the evolved distributions as functions of the parton momentum fraction $x$,
not their Mellin transforms as functions of the conjugate variable $z$.  To
obtain the evolved parton distribution in $x$-space, we must invert the
Mellin transform,
$$
f(x,Q^2) = {1\over 2\pi i}\int_{C} dz\; x^{-z} f^z(Q^2)\,,
\anoneqn$$
where the contour $C$ runs parallel to the imaginary axis, to the right of
the rightmost pole in the integrand.

Up to this point, everything I have written is completely standard.  I
now wish to investigate how one may deform the contour $C$ in order
to obtain an efficient method for evaluating the integral.  Note that the
integrand has poles only along the real axis; denote the rightmost pole
by $c_r$.  (It is typically around $0.5$ for the non-singlet case, and
$1.3$ for the singlet case.)

One can in principle perform the contour integral along the textbook
contour $z=c+i y$ ($-\infty<y<\infty$ and $c>c_r$),
as was done in the original work of Gl\"uck and Reya~[\use\GR].
One then obtains the expression
$$\eqalign{
{x^{-c}\over2 \pi}&\int_0^{\infty} dy\;
\L x^{-i y} f^{c+i y}(Q^2) + x^{i y} f^{c-i y}(Q^2)\R
\cr &={x^{-c}\over\pi}\int_0^{\infty} dy\;
{\rm Re}\LB x^{-i y} f^{c+i y}(Q^2) + x^{i y} f^{c-i y}(Q^2)\RB\,.
}\anoneqn$$

The large-$y$ behavior of the integrand is determined by the
behavior of the initial distribution as $x\rightarrow 1$.
Initial distributions of the form $x^a (1-x)^\beta$ give
rise to a power decay, $f \sim y^{-\beta-1}$ as $y\rightarrow\infty$.
Because the contour is parallel to the imaginary axis, the $x^{-z}$
factor is a purely oscillating; it does not damp the integrand as
$y\rightarrow\infty$.  On the other hand, because the integrand has
no poles off the real axis, and because the integral along a half-circle
at $\infty$ in the left half-plane (or along part of this half-circle)
vanishes, we can
freely deform the contour into the left-half plane, so
long as we stay away from the real
axis.  Were we to choose a contour such that $z$ has an increasingly
negative real part as it heads off to infinity, the $x^{-z}$ factor
would contribute an exponential suppression, improving the convergence
of the integral.  (Recall that the base point $x$ lies between $0$ and $1$.)

This motivates the choice of contour in ref.~[\use\GRV], where it
is taken to contain two line segments running
diagonally into the left-half plane from a point $c$ on
the real axis to infinity.

However, we might ask: why {\it this\/} contour?  Or, phrased differently: how
{\it should\/} we choose a contour?  I address this question in
the next section.

\section{Choosing a Contour}
\tagsection\ContourChoiceSection
\vskip 10pt

The most obvious answer is that we should choose the contour of
steepest descent; this choice will yield an integral that converges
most efficiently and (one therefore hopes) can be evaluated numerically with
fewest function evaluations for a fixed desired accuracy.
One might fear that finding the contour of steepest descent requires
a complicated computation.  For our problem, however, it will
turn out that there is a simple but very good approximation to the desired
contour.

Now, we don't want to find a contour for each value of $Q^2$, especially
if finding such a contour involves a significant number of function
evaluations; this would defeat the whole purpose of the exercise!
For use in a program evaluating a cross section, it is convenient
to set up a grid of $x$ and $Q^2$ points ({\it \`a la\/} MRS [\use\MRS]),
and interpolate between them.
Instead of finding a contour for each $(x,Q^2)$ pair,
use the following strategy: for each grid $x$ value,
find the contour of steepest descent for $Q_0^2$, and then use it
for evolution to all $Q^2$ values.  Since the evolution is relatively
slow (it is only in $\ln\ln Q^2$, after all),
the contour at $Q_0^2$ should be fairly close to the contour of
steepest descent at $Q^2$.   From a programming point of view,
this approach allows all the contours, points
along them, and the associated anomalous dimensions to be computed as
setup code; only the evolution operators at the given points will need
to be evaluated anew for each $Q^2$ to which we wish to evolve the
distributions.

Let us then first consider the inverse
Mellin transform of $f^z(Q_0^2)$.  (Yes, we already know the answer ---
it is just $f(x,Q_0^2)$ --- but that is not the point.)
There are certain technical complications
in the singlet sector, which I will discuss section~\use\SingletEvolution,
so let us restrict attention here to the non-singlet distributions.

The first step in determining the contour is to find the minimum of the
integrand along the real axis.  As the starting distribution is
a positive function, its Mellin transform will be positive along the
real axis to the right of its rightmost pole $c_r$.  Furthermore, the starting
distribution is not infinitely concentrated at $x=1$, so its Mellin
transform will decrease as $z\rightarrow\infty$.  On the other hand,
since $x<1$,
$x^{-z}$ increases exponentially as $z\rightarrow \infty$.  The product
of the two must therefore have a minimum somewhere, and the value of
the product there will be positive.
(If there happen to be multiple minima,
pick the one closest to the rightmost pole.)
Call this point $c_0$.

The integrand is analytic as a function of $z$. The minimum of
the function along the real axis is therefore actually
a saddle point of the function in the complex plane,
and thus the place to start tracing
out our desired contour.  Furthermore (again because of analyticity),
the desired contour is also a contour of stationary phase, so that
the integrand will be real along it. (I will also assume that the rightmost
singularity of the Mellin transform of the initial distribution is
to the right of the rightmost pole in the anomalous dimensions; this
assumption, which holds for realistic parton distributions, ensures
that the minimum is only slightly different for $f^z(Q^2)$ than for
$f^z(Q_0^2)$, and is in any event necessary if we are to use a contour as
determined
below for integrating $f^z(Q^2)$.)

\def\Csb{{\overline C}_s}
Define $F(z) = x^{-z} f^z (Q_0^2)$;
our inverse Mellin transform then has the form
$$\eqalign{
I &= {1\over 2\pi i}\int_{C_s+(-\Csb)} dz\;
   x^{-z} f^z(Q_0^2)
={1\over 2\pi i}\int_{C_s} \LB dz\; F(z) - d\bar z\; F(\bar z)\RB\cr
&={1\over \pi} \int_{C_s} \Re\LB -i\, dz\; F(z)\RB\cr
}\anoneqn$$
where $C_s$ is the part of the
contour of steepest descent running upwards from $c_0$.

Let us examine the contour near the point $c_0$.  A generic saddle point
will have a non-vanishing second derivative (and all the relevant minima
for conventional choices of $f(x,Q_0^2)$ are indeed generic), so the
contour will start out with a tangent parallel to the imaginary axis:
that is the direction of steepest descent of
$$
F(z) \sim F(c_0) + {F''(c_0)\over2}\,(z-c_0)^2
  + \cdots\,,
\eqn\Expansion$$
which is perhaps easiest to see if we parametrize $z(t) = x(t) + i y(t)$,
with
$$
x(t_0=0) = c_0,\qquad y(0) = 0,\qquad
x,y {\rm\ real}\,.
\anoneqn$$
Note that the symmetry of the contour under reflection in the
real axis forces $x$ to
be an even function of $t$ (so that $x'(0)=0$),
and $y$ to be an odd function. We can
rescale $t$ to make $y'(0) = 1$.
  The expansion in eqn.~(\use\Expansion) is then
$$\eqalign{
F(z(t)) &\sim F(c_0)
   + {F''(c_0)\over2} \, (x'(0)^2-y'(0)^2 +2i x'(0) y'(0))\,t^2
   + \cdots\cr
 &=  F(c_0)
   - {F''(c_0)\over2} \, t^2
   + \cdots\cr
}\anoneqn$$
Since all the derivatives of $F(x)$ are real, the equation
$\Im F(z(t)) = 0$ is then satisfied to this order, and as
$F''(c_0)$ is positive, the function decreases with $t$.
However, for conventional choices of $f(x,Q_0^2)$, the contour does
not continue parallel to the imaginary axis; to see where it does go,
we need to consider the expansion to one higher order,
$$
F(z(t)) \sim F(c_0)
   - {F''(c_0)\over2} \, t^2
  + {1\over 6}\L -i F^{(3)}(c_0)+3i F''(c_0) x''(0)\R\,t^3
  + \cdots
\anoneqn$$
To ${\cal O}(t^3)$, $\Im F(z(t)) = 0$ then requires
$$
x''(0) = {F^{(3)}(c_0)\over3 F''(c_0)}
\eqn\Xdderivative$$
Thus in the neighborhood of $c_0$, the contour has the form
$$ z(t) = c_0 + i t + {F^{(3)}(c_0)\over6 F''(c_0)} t^2\,.
\eqn\Contour$$
($F^{(3)}(c_0)$ is typically negative for the class of functions
in which we are interested, as is necessary for this to be a useful
truncation.)
What is not so obvious --- but turns out to be true --- is that for
our purposes, the contour $C'$ described by this equation is
essentially indistinguishable from the true contour of steepest
descent.  That is, in the region where the integrands of interest give
the bulk of the contributions to the inverse Mellin transform,
the two contours are extremely close, and so we incur almost
no efficiency penalty in choosing the simplified contour given
by eqn.~(\use\Contour).  With this choice, our inverse Mellin
transform can now be written
$$
{1\over \pi} \int_0^\infty dt\; \Re\LB
 \L 1-i {F^{(3)}(c_0)\over3 F''(c_0)} t\R\; F(z(t))\RB\,.
\eqn\Tintegral$$

As we shall see in later sections, this quadratic contour is a significant
improvement over the linear contour chosen in ref.~[\use\GRV], not to mention
the textbook contour.

\section{Evaluating the Inverse Transform}
\tagsection\InverseTransformSection
\vskip 10pt

The next point we must consider is the evaluation of the inverse Mellin
transform using our new contour~(\use\Contour).  We want to choose a few
points along the contour at which to evaluate the function in order to
approximate the integral by a finite sum,
$$
{1\over \pi} \int_{C'} \Re\LB -i dz\; F(z)\RB
   \simeq \sum_{i| z_i\in C'} w_i F(z_i)
\anoneqn$$

In order to find a `good' set of points, we should
in principle first find a set
of functions in which to expand $f^z(Q^2)$.  A possible
(though not necessarily ``optimal'') choice is a set of orthogonal
polynomials with an astutely-chosen weight function.  In fact, for
integrating $f^z(Q_0^2)$,
such a choice {\it is\/} optimal if we pick the weight
function to be $F(z)$ itself!  We might therefore
be tempted to construct
a set of orthogonal polynomials with respect to this weight function.

First, though, let us examine the behavior of the integrand along the
contour near $c_0$.
Assume that $f^z(Q_0^2)$ does not vary exponentially, so that
it cannot eliminate the exponential fall-off expected from
the $x^{-z}$ factor.  We then expect a behavior of the form
$F(c_0) e^{-g(t)}$, where the power series expansion of $g(t)$ can
be determined by matching coefficients with the power
series expansion in the integrand of eqn.~(\use\Tintegral).
(Note that the imaginary part of $z'(t)$ in the prefactor does not contribute
until order $t^5$.)  We then obtain
$$
g(t) \sim {F''(c_0)\over2 F(c_0)} \, t^2 + \cdots
\anoneqn$$
(As discussed in the previous section, both $F''(c_0)$ and $F(c_0)$
will be positive.)
This suggests that we perform a change of variables
$u = {F''(c_0)\over2 F(c_0)} \, t^2$, whereupon the integral of
eqn.~(\use\Tintegral) becomes
$$\eqalign{
{1\over \pi}\sqrt{{F(c_0)\over2 F''(c_0)}}
  &\int_0^\infty {du\over\sqrt{u}}\; \Re\LB
 \L 1-i {F^{(3)}(c_0)\over3 F''(c_0)} \sqrt{2 F(c_0)\over F''(c_0)}
\sqrt{u}\R\;
   F\L z\L \sqrt{2 F(c_0)\over F''(c_0)}\sqrt{u}\R\R\RB\cr
&={c_2\over 2\pi} \int_0^\infty {du\over\sqrt{u}}e^{-u}\; \Re\LB
 e^u \L 1-i c_2 c_3\sqrt{u}\R\;
   F\L z\L c_2\sqrt{u}\R\R\RB\cr
}\eqn\Uintegral$$
where $c_2 = \sqrt{2 F(c_0)/F''(c_0)}$ and $c_3 = F^{(3)}(c_0)/(3  F''(c_0))$.
For small $u$, we expect the factor inside the brackets on
the second line to vary slowly.
What is again not so obvious, but turns out to be true for
initial parton distributions of interest, is that the factor in
brackets varies {\it very\/} little, and smoothly at that,
 in the entire region where the integral
receives noticeable contributions.  It is thus an excellent candidate
for approximation by polynomials.  The same statements continue to hold
for the inverse Mellin transform of $f^z(Q^2)$, in which the factor inside
the brackets in eqn.~(\use\Uintegral) is now multiplied by the evolution
operator $E^z(a_s(Q^2),a_s(Q_0^2))$.

The reader may also recognize the prefactor in front of the brackets
as the weight function for the generalized Laguerre polynomials
$L^{(-1/2)}_n(u)$, and it is a generalized
Gauss-Laguerre quadrature formula employing these polynomials which we
should use for evaluating the integral in eqn.~(\use\Uintegral).  This
formula approximates an integral
$$
\int_0^\infty {du\over\sqrt{u}}e^{-u}\; h(u) \simeq \sum_{j=1}^n w_j h(u_j)\,,
\anoneqn$$
where the $u_j$ are the zeros of $L^{(-1/2)}_n(u)$, and the weights are
given by standard formul\ae~[\use\DR],
$$
w_j = {\Gamma(n+1/2)\over n!\, (n+1)^2}
   {u_j\over \LB L^{(-1/2)}_{n+1}(u_j)\RB^2}\,.
\anoneqn$$

\section{Singlet Evolution}
\tagsection\SingletEvolution
\vskip 10pt

The same considerations discussed in the previous two sections
apply to singlet evolution, with some
important differences.   In the non-singlet case, the integrand
in the inverse Mellin transform of $f^z(Q_0^2)$ is modified by a simple
(and modest) multiplicative factor to obtain the integrand for $f(x,Q^2)$.
In contrast, in the singlet case, the evolution operator has non-trivial
matrix structure, and is {\it not\/} close to the identity matrix even
for $Q^2$ near $Q_0^2$.  This happens because the $(\Sigma,g)$ basis is
not an eigenbasis even for the lowest-order Altarelli-Parisi function.
Accordingly, contours chosen
according to either $\Sigma^z$ or $g^z$ will not be particularly
good ones.  What we want is a basis in which the evolution operator
does not twist one direction into another as we move around in the
complex plane.  Such a basis is given
by the eigenvectors of the lowest-order Altarelli-Parisi
matrix, that is by $P_{\pm}^z ({\Sigma\atop g})$.
Thus in the singlet sector, we could rewrite our Mellin integral as
\def\svecE{\L{\Sigma^z(Q^2)\atop g^z(Q^2)}\R}
\def\svec{{\bf s}^z(Q^2)}
\def\svecz{{\bf s}^z(Q_0^2)}
$$\eqalign{
&\int_{C} \Re\LB -i dz\; x^{-z} E^z(a_s,a_0) \svecz\RB
=\cr
&\hskip 15mm \int_C \Re\LB -i dz\; x^{-z} E^z(a_s,a_0) P_{-}^z\svecz\RB
+ \int_C \Re\LB -i dz\; x^{-z} E^z(a_s,a_0) P_{+}^z\svecz\RB\,,
}\anoneqn$$
where
$$
\svec = \svecE\,.
\anoneqn$$

In each of these integrals, each of the components is now modified
multiplicatively (up to ${\cal O}(a_s)$ corrections), like the
non-singlet Mellin integrals I analyzed in previous sections.
We might expect them to be treatable in the same fashion --- contours chosen
according to the components of $P_{\pm}^z\svecz$.  Indeed, following
this approach we would choose four different contours for the
different integrations above.  There are, however, two complications
we would confront.

Analyticity assures us that it is legitimate to choose different contours
for the $P_{-}^z$ and $P_{+}^z$ integrals, but there is a subtlety: the
integrands have branch cuts in the complex plane, and the contours chosen
according to section~\use\ContourChoiceSection\ may cross these branch
cuts.  So long as we use the same contour for both integrals, this
is completely harmless, because all that happens upon crossing a branch
cut is that $\lambda_{\pm}^z$ interchange roles.  This merely interchanges
the two integrands, and the sum is unaffected.
If we want to choose different contours, however, we should either shift
the branch cuts so that neither integral crosses them (which is possible
only some of the time), or else we must compute the integral around the branch
cut.
While the latter is possible, it will not allow an efficient evaluation of the
integral.

The other complication is that the components of $P_{\pm}^z\svecz$ are
no longer necessarily positive functions. As a result, they no longer
necessarily have a minimum to the right of the rightmost pole.  (Typically
$\Sigma^z$ and $g^z$ will have nearly the same
rightmost pole, as it controls the $x\rightarrow 0$ behavior;
the gluon distribution
will be slightly steeper [\use\MRS].)  Indeed,
in practice one combination --- $P_{-}^z\svecz$ --- is roughly a sum of
the two components, and hence is positive, and can be handled according
to the prescriptions of the previous section; while the other combination
($P_{+}\svec$) is not, and in fact has no minimum to the right of
the rightmost pole for certain choices of $f^z(Q_0^2)$.
(The contour of steepest descent heads straight into the pole.)

Thus instead it is better to pick a slightly different approach to choosing the
contour-determining function in this case.
Instead of the parton distributions at the original
scale $Q_0^2$, take the distributions evolved (using leading-order evolution)
to another
 fixed scale $Q_1^2$ as the contour-determining function $F(z)$.  The latter is
of course just given by the evolution operator $E^z_{\rm LO}(Q_1^2,Q_0^2)$
multiplied
by the starting distributions $\svecz$.  While in principle we might choose
different
contours for the two different components in the singlet sector, in practice it
is more efficient overall to choose a single contour using the gluon component
 in the role of $F(z)$.  That is, for the purposes of determining the contour
of integration, and the integration points along it, take
$$
\LB \L P^z_{-}\,\L{a_s(Q_1^2)\over a_0}\R^{-\lambda_{-}^z/\beta_0}
+P^z_{+}\,\L{a_s(Q_1^2)\over a_0}\R^{-\lambda_{+}^z/\beta_0}\R\,\svecz\RB_{2}
\anoneqn$$
for $F(z)$ in the singlet sector.  In this expression,
 $Q_1$ would be a scale intermediate between
$Q_0^2$ and the highest scale to which we wish to evolve the parton
distributions.

\section{Numerical Performance}
\tagsection\NumericalPerformance
\vskip 10pt

\def\MSbar{{\overline {\rm MS}}}
At how many points do we need to evaluate the integrands constructed
according to the prescriptions in previous sections, in order to obtain
an evolved parton distribution function to a given accuracy?  While the
answer will depend on the precise form of the initial distribution
functions, and on the desired accuracy, we can obtain a very good idea
by studying the numerics of the evolution of the toy
parton distribution set considered by Bl\"umlein et al.~[\use\Compare].
Those authors consider the $\MSbar$ massless four-flavor
evolution from a reference scale $Q_0 = 2$~GeV,
with $\Lambda_{\MSbar}^{(4)} = 250$~MeV.  The initial distributions are
taken to have the following form,
$$\eqalign{
u_v(x,Q_0^2) &= A_u x^{-0.5} (1-x)^3,\hskip 10mm
d_v(x,Q_0^2) = A_d x^{-0.5} (1-x)^4,\cr
S(x,Q_0^2) &= A_S x^{-1.2} (1-x)^7,\hskip 10mm
c(x,Q_0^2) = 0,\cr
g(x,Q_0^2) &= A_g x^{-1.2} (1-x)^5\,.
}\anoneqn$$
The sea $S(x,Q^2) = \LB \Sigma-u_v-d_v\RB(x,Q^2)$ includes the charm
content, and is taken to carry $15\%$ of the nucleon momentum at the
input scale.  This fixes $A_S$; the remainining $A_i$ are fixed by
the flavor and momentum sum rules.

Take the desired accuracy to be the  same $2$ parts in $10^{4}$ sought
by the above authors.  For the non-singlet evolution (for example, the
valence  up or down densities) to  $Q=10$~GeV requires  around 30
points  at  small  $x$, rising  to   around 80   points   for $x=0.7$,   using
Gauss-Legendre   quadrature  on    the  contour of    ref.~[\use\GRV].
(Evolution to $Q=10$~TeV requires about   10\% more points at   larger
$x$.)    In  constrast, using   contour  derived  in  the present work
requires only three  or four points for  all $10^{-5} <  x < 0.9$.  In
the  singlet sector, the new contour   typically requires four points,
while the  contour  of  ref.~[\use\GRV] requires  roughly the same
number of points as the non-singlet sector at  small $x$, and somewhat
more at larger  $x$.   Obtaining the toy parton set to the stated
accuracy typically requires more points, because the charm distribution
is a small difference of two larger numbers.
The quadratic contour derived in this paper requires a lone
additional
point for most $x$ values, and a total of seven points at the largest $x$
value.  (For the contour of ref.~[\use\GRV], the toy parton set typically
requires 30 points at small $x$ to over 120 at the largest $x$ value.)

The answers  obtained  using this new  approach
should be  compared with those  of  the `truncated analytic solution,'
that is the lower half, of the table  of ref.~[\use\Compare].  The
following table illustrates the convergence of sample values using
the new contour, with one, two, or four Gauss-Laguerre points,
\def\fracerr{{\rm frac.\ error}}
\def\vspan#1#2{\parindent0pt\setbox0\vbox to #1\normalbaselineskip
 {\null\vfill#2\vfill\null}%
 \ht0\ht\strutbox\dp0\dp\strutbox
 \setbox1\hbox{#2}\wd0\wd1\box0}
$$
\vbox{\offinterlineskip
\hrule
\halign{%
&\vrule# & \strut\quad\hfil$#$\quad\cr
height2pt&\omit&&\omit&&\omit&\omit&\omit&\omit&\omit&\omit&\omit&\cr
& \vspan2{$x$}\hfil && \vspan2{$f$}\hfil && \multispan7 \hfil{$x\,
f(x)$}\hfil&\cr
height2pt&\omit&&\omit&&\omit&\omit&\omit&\omit&\omit&\omit&\omit&\cr
&\omit &\omit& \omit&& \multispan7 {\hskip -3pt\hrulefill}\cr
height2pt&\omit&&\omit&&\omit&&\omit&&\omit&&\omit&\cr
&  &&  &&  n=1\hfil && 2\hfil && 4\hfil && {\rm exact}\hfil &\cr
height2pt&\omit&&\omit&&\omit&&\omit&&\omit&&\omit&\cr
\noalign{\hrule}
height2pt&\omit&&\omit&&\omit&&\omit&&\omit&&\omit&\cr
& 10^{-5} && u_v\hfil && 0.0096683 && 0.0094433 && 0.0094111
                    && 0.0094114&\cr
& && \fracerr && 2.7 \cdot 10^{-2}&& 3.4 \cdot 10^{-3}&& -2.7 \cdot 10^{-5}
                    && &\cr
height2pt&\omit&&\omit&&\omit&&\omit&&\omit&&\omit&\cr
\noalign{\hrule}
height2pt&\omit&&\omit&&\omit&&\omit&&\omit&&\omit&\cr
& 10^{-3}&& u_v\hfil && 0.089636 && 0.088177 && 0.088077
                    && 0.088086&\cr
& && \fracerr && 1.8 \cdot 10^{-2}&& 1.0 \cdot 10^{-3}&& -1.0 \cdot 10^{-4}&&
                   &\cr
height2pt&\omit&&\omit&&\omit&&\omit&&\omit&&\omit&\cr
\noalign{\hrule}
height2pt&\omit&&\omit&&\omit&&\omit&&\omit&&\omit&\cr
& 0.1&& u_v\hfil && 0.47445 && 0.47291 && 0.47267
                    && 0.47267&\cr
& && \fracerr && 3.8 \cdot 10^{-3}&& 5.2 \cdot 10^{-4}&& 7.4 \cdot 10^{-7} &&
&\cr
height2pt&\omit&&\omit&&\omit&&\omit&&\omit&&\omit&\cr
\noalign{\hrule}
height2pt&\omit&&\omit&&\omit&&\omit&&\omit&&\omit&\cr
& 0.3&& u_v\hfil && 0.31061 && 0.30793 && 0.30797
                    && 0.30797&\cr
& && \fracerr && 8.6 \cdot 10^{-3} && -1.2 \cdot 10^{-4}&& 4.1 \cdot 10^{-6}&&
                      &\cr
height2pt&\omit&&\omit&&\omit&&\omit&&\omit&&\omit&\cr
\noalign{\hrule}
height2pt&\omit&&\omit&&\omit&&\omit&&\omit&&\omit&\cr
& 10^{-5}&& g\hfil && 102.92 && 97.879 && 98.080
                    && 98.080&\cr
& && \fracerr && 4.9 \cdot 10^{-2}&& -2.0 \cdot 10^{-3}&& 4.4\cdot 10^{-7} &&
&\cr
height2pt&\omit&&\omit&&\omit&&\omit&&\omit&&\omit&\cr
\noalign{\hrule}
height2pt&\omit&&\omit&&\omit&&\omit&&\omit&&\omit&\cr
& 10^{-3} && g\hfil && 22.052 && 21.106 && 21.112
                    && 21.112&\cr
& && \fracerr && 4.5\cdot 10^{-2} && -3.2 \cdot 10^{-4}&& 6.8\cdot 10^{-6}&&
&\cr
height2pt&\omit&&\omit&&\omit&&\omit&&\omit&&\omit&\cr
\noalign{\hrule}
height2pt&\omit&&\omit&&\omit&&\omit&&\omit&&\omit&\cr
& 0.1&& g\hfil && 1.4168 && 1.4152 && 1.4151
                    && 1.4151 &\cr
& && \fracerr && 1.2 \cdot 10^{-3}&& 1.5 \cdot 10^{-5} && 1.5\cdot 10^{-7} &&
&\cr
height2pt&\omit&&\omit&&\omit&&\omit&&\omit&&\omit&\cr
\noalign{\hrule}
height2pt&\omit&&\omit&&\omit&&\omit&&\omit&&\omit&\cr
& 0.3&& g\hfil && 0.18849 && 0.18754 && 0.18755
                    && 0.18755&\cr
& && \fracerr && 5.0\cdot 10^{-3} && -3.8 \cdot 10^{-5}&& -2.2\cdot 10^{-7}&&
&\cr
height2pt&\omit&&\omit&&\omit&&\omit&&\omit&&\omit&\cr}
\hrule}
$$

\section{Crossing Functions}
\tagsection\CrossingFunctions
\vskip 10pt

\def\smin{s_{\rm min}}
\def\qb{{\overline q}}
Crossing functions arise naturally in a crossing-symmetric
formalism~[\use\GGK] for evaluating general next-to-leading order
distributions in a collider environment.  They give the change in
the differential cross section as a colored final-state particle is crossed
from the final state into the initial state.
The crossing functions $C_{a\leftarrow p}(x,Q^2)$
are factorization-scheme dependent.
They can be expressed in terms of scheme-independent functions
$A_{a\leftarrow p}$
and scheme-dependent functions $B_{a\leftarrow p}$ as follows,
$$
C_{a\leftarrow p}(x,Q^2) = \left({N\over2\pi}\right)
\left[ A_{a\leftarrow p}(x,Q^2)\ln\left({\smin\over Q^2}\right)
+ B_{a\leftarrow p}(x,Q^2)\right]\,,
\anoneqn$$
where $\smin$ is the boundary between the real contributions integrated
analytically and those integrated numerically.  A physical quantity
will be independent of $\smin$, in the limit that $\smin\rightarrow 0$.
The $A$ and $B$ functions are convolutions of universal kernels with
the parton distribution functions,
$$\eqalign{
A_{a\leftarrow p}(x,Q^2) &= K^A_{a\leftarrow b}(x) \otimes f_{b\leftarrow
p}(x,Q^2)\,,\cr
B_{a\leftarrow p}(x,Q^2) &= K^B_{a\leftarrow b}(x) \otimes f_{b\leftarrow
p}(x,Q^2)\,,\cr
}\anoneqn$$
\def\frac#1#2{{#1\over#2}}
(again with implicit summation over $b$).  Explicit expressions for the kernels
$K^{A,B}$ may be extracted from ref.~[\use\GGK],
$$\hskip-5pt\eqalign{
K^A_{g\leftarrow g}(x) &=
  \frac{(11N-2n_f)}{6N} \delta(1-x)
 + 2\left( \frac{x}{(1-x)_+}+ \frac{(1-x)}{x}+ x(1-x) \right)\,,\cr
K^A_{q\leftarrow q'}(x) &= \delta_{qq'}
\left(1-\frac{1}{N^2}\right)
\LB\frac{3}{4}\delta(1-x)
+ \frac{1}{2}\left(\frac{1+x^2}{(1-x)_+}\right)\RB\,,\cr
K^A_{q\leftarrow g}(x)
 &= \frac{1}{2 N}\L z^2 + (1-z)^2\R\,,\cr
K^A_{g\leftarrow q}(x)
&= \frac{1}{2}\L1-{1\over N^2}\R\,{1+(1-z)^2\over z}\,,\cr
K^{B,\MSbar}_{g\leftarrow g}(x) &=
\left(\frac{\pi^2}{3}-\frac{67}{18}+\frac{5n_f}{9N}\right) \delta(1-x)
  + 2 x\left(\frac{\ln(1-x)}{(1-x) }\right)_+
\cr&\hskip 30mm
  +2\left(\frac{(1-x)}{x}+ x(1-x) \right)\ln(1-x)
  \,,\cr
K^{B,\MSbar}_{q\leftarrow q'}(x) &=
 \delta_{qq'}\left(1-\frac{1}{N^2}\right)\,\LB
\left(\frac{\pi^2}{6}-\frac{7}{4}\right) \delta(1-x)
-\frac{1}{2} (1-x) +
\frac{1}{2}(1+x^2)\left(\frac{\ln(1-x)}{(1-x)}\right)_+ \RB\,,\cr
K^{B,\MSbar}_{q\leftarrow g}(x)
&= {1\over 2N} \LB \L x^2 + (1-x)^2\R\,\ln(1-x)+1-x^2-(1-x)^2\RB\,,\cr
K^{B,\MSbar}_{g\leftarrow q}(x)
&= {1\over2}\L1-{1\over N^2}\R\,\LB {1+(1-x)^2\over x}\,\ln(1-x)+x\RB\,.\cr
}\eqn\Kernels$$
(Note that $K^{A,B}_{\qb\leftarrow \qb'} = K^{A,B}_{q\leftarrow q'}$, etc.)
In this equation the $(~)_+$ prescription is defined by
$$
\left(F(z)\right)_+ = \lim_{\beta \to 0} \left(
\theta(1-z-\beta) F(z)-\delta(1-z-\beta)\int^{1-\beta}_0 F(y) \, dy \right),
\anoneqn$$
such that,
$$\eqalign{
\int^1_x dz  \frac{g(z)}{(1-z)_+} &=
\int^1_x dz \frac{g(z)-g(1)}{1-z}+g(1)\log(1-x)\,,\cr
\int^1_x dz g(z)\left(\frac{\log(1-z)}{1-z}\right)_+ &=
\int^1_x dz \frac{g(z)-g(1)}{1-z}\log(1-z) +\frac{g(1)}{2}\log^2(1-x)\,,\cr
}\anoneqn$$
provided that $g(z)$ is a function well behaved at $z=1$.

The Mellin moments of the kernels are
$$\eqalign{
K^{A,z}_{g\leftarrow g} &=
  {11\over 6}-{n_f\over 3N}
   + 2\left( \psi(1)-\psi(z+1)
           +{1\over z(z-1)}+ {1\over (z+1)(z+2)}\right)\,,\cr
K^{A,z}_{q\leftarrow q'}
&= \delta_{qq'} \left(1-\frac{1}{N^2}\right)
  \LB\frac{3}{4}+ \psi(1)-\psi(z+1)+{1\over 2 z (z+1)}\RB\,,\cr
K^{A,z}_{q\leftarrow g} &=
   \frac{1}{2 N} \L {1\over z+2}+ {2\over z (z+1) (z+2)}\R\,,\cr
K^{A,z}_{g\leftarrow q} &=
  \frac{1}{2}\L1-{1\over N^2}\R\,
    \L {1\over z-1} + {2\over z (z-1) (z+1)}\R\,,\cr
K^{B,\MSbar,z}_{g\leftarrow g} &=
\frac{\pi^2}{3}-\frac{67}{18}+\frac{5n_f}{9N}
+ (\psi(z+1)-\psi(1))^2+\psi'(1)-\psi'(z+1)
\cr&\hskip 5mm +\L{2\over z(z-1)}+{2\over (z+1)(z+2)}\R\,
\L 1+\psi(1)-\psi(z+1)\R-{2\over (z+1)^2}+{2\over (z+2)^2}\,,\cr
K^{B,\MSbar,z}_{q\leftarrow q'} &=
\delta_{qq'}\left(1-\frac{1}{N^2}\right)\,\LB
\frac{\pi^2}{6}-\frac{7}{4}
  +{1\over 2z^2}-{1\over 2 z (z+1)^2}
  +{1\over 2} (\psi(z+1)-\psi(1))^2 \RP\cr
&\hskip 20mm\LP
  +{1\over 2 z(z+1)} (\psi(1)-\psi(z+1))
  +{1\over2}(\psi'(1)-\psi'(z))\RB\,, \cr
K^{B,\MSbar}_{q\leftarrow g} &=
  {1\over 2N} \LB \L {1\over z+2}+{2\over z(z+1)(z+2)}\R\,(\psi(1)-\psi(z+3))
 \RP\cr &\hskip 40mm\LP
                +{2\over z(z+2)}+{1\over z(z+1)(z+2)}\RB\,,\cr
K^{B,\MSbar,z}_{g\leftarrow q} &=
   {1\over2} \L 1- {1\over N^2} \R\LB
   \L  {1\over z-1} + {2\over z(z-1)(z+1)} \R\L \psi(1)-\psi(z+1)\R
    + {z\over (z+1)^2} + {2\over z(z-1) }\RB\,.\cr
}\eqn\KernelMoments$$

These moments allow us to write
$$\eqalign{
A_{a\leftarrow p}^z(Q^2) &=  K^{A,z}_{a\leftarrow b}\,f_{b\leftarrow p}^z(Q^2)
\cr
B_{a\leftarrow p}^z(Q^2) &=  K^{B,z}_{a\leftarrow b}\,f_{b\leftarrow p}^z(Q^2)
}\anoneqn$$
and then using eqn.~(\use\EvolutionOperatorDef),
$$\eqalign{
A_{a\leftarrow p}^z(Q^2) &=  K^{A,z}_{a\leftarrow b}\,
E^z_{bc}(\alpha_s(Q^2),\alpha_0) f_{c\leftarrow p}^z(Q_0^2)
\cr
B_{a\leftarrow p}^z(Q^2) &=  K^{B,z}_{a\leftarrow b}\,
E^z_{bc}(\alpha_s(Q^2),\alpha_0) f_{c\leftarrow p}^z(Q_0^2)
\cr}\eqn\CrossingMoment$$
We can evaluate the crossing function, given by the inverse Mellin
transform of this moment, using the same approach described in
previous sections, but with $f^z(Q_0^2)$ replaced by $K^{X,z} f^z(Q_0^2)$
($X=A, B$).

The conjugation identities mentioned above, along with the fact
that $K^{X,z}_{q\leftarrow \qb'}=0$, imply that the non-singlet and
singlet sectors do not mix in eqn.~(\use\CrossingMoment), and that
the kernel in the non-singlet sector is simply $K^{X,z}_{q\leftarrow q}$,
while the $2\times 2$ matrix in the singlet sector is
$$
\L\matrix{K^{X,z}_{q\leftarrow q}& K^{X,z}_{q\leftarrow g}\cr
          K^{X,z}_{g\leftarrow q}& K^{X,z}_{g\leftarrow g}\cr}\R\,.
\anoneqn$$

\section{Comparison with $x$-Space Codes}
\tagsection\XSpaceSection
\vskip 10pt

As mentioned in the introduction, the two main methods now commonly
used to evolve parton distribution functions are
the Mellin transform method~[\use\GRV,\use\Mellin] pursued in
previous sections, or direct integration~[\use\Direct] of the differential
equations for $f(x,Q^2)$.  While formally equivalent to NLO,
these techniques differ in the higher-order terms implicitly retained.
As noted by Bl\"umlein et al.~[\use\Blum], these differences can
have a substantial effect on the parton distributions, as much as several
percent even at moderate $x$.  One lesson from this comparison is that
a determination of $\alpha_s(M_Z^2)$ to 1\% will require NNLO calculations;
in the meantime, however, it is useful to be able to reproduce answers
obtained through the $x$-space method as well.  The purpose of this section
is to show that solutions identical to those obtained by direct integration
can be obtained using the Mellin transform technique, taking advantage
of the same enhancements considered in previous sections.

Let us first consider $a_s$ itself; in terms of the QCD scale
parameter $\Lambda$, it can be written~[\use\FP] at NLO as
$$
a_s(Q^2) = {1\over\beta_0 \ln Q^2/\Lambda^2} \LB
      1-{\beta_1\over\beta_0^2} {\ln\ln Q^2/\Lambda^2\over\ln Q^2/\Lambda^2}
                +{\cal O}\L \ln^{-3} Q^2/\Lambda^2\R\RB\,;
\eqn\ApproximateAlphaS$$
let us replace it with an
approximation (to NLO) for $a_s$ in terms of
$a_s$ at a different scale, instead of one in terms of $\Lambda_{\rm QCD}$,
$$\eqalign{
a_s(Q^2,Q_0^2,a_0) &=
  {a_0 (1 + a_0 \beta_0 \ln Q^2/Q_0^2)\over
    \L1 + a_0 \beta_0 \ln Q^2/Q_0^2\R^2 +
     a_0 \beta_r \L 1 + a_0 \beta_r + a_0 \beta_0 \ln Q^2/Q_0^2\R\,
     \ln \LB 1 + {a_0 \beta_0 \ln Q^2/Q_0^2\over
                 (1 + a_0 \beta_r)}\RB}
   \,,\cr
}\eqn\NewAlphaS$$
where $\beta_r \equiv \beta_1/\beta_0$, and
the number of flavors is taken to be constant throughout the interval
$[Q_0,Q]$.  This particular form of $a_s$ emerges by performing {\it one\/}
Newton-Raphson improvement to the 1-loop solution.
We could recover the solution of the implicit (beta function)
 equation for $a_s$,
$$
{1\over a_s(Q^2)} = {1\over a_s(Q_0^2)} + \beta_0 \ln \L Q^2/Q_0^2\R
-{\beta_1\over\beta_0} \ln\LB{ a_s(Q^2) (\beta_0+\beta_1 a_s(Q_0^2)
                               \over a_s(Q_0^2) (\beta_0+\beta_1 a_s(Q^2)}\RB
\eqn\Implicit$$
\def\asu{a_s^{\rm u}}
if we iterate the evolution\footnote{$^{\dagger}$}{Iteration will
result in convergence to the numerical solution of eqn.~(\use\BetaFunction)
only if the form of the approximation is suitable; for a slowly varying
function whose true zero is near the initial approximation, a
Newton-Raphson approximation is indeed suitable.},
$$\eqalign{
a_s^{[n]}(Q^2,Q_0^2,a_0) &=
        a_s^{[n-1]}(Q^2,Q_1^2,a_s^{[n-1]}(Q_1^2,Q_0^2,a_0))\,,\cr
a_s^{[0]}(Q^2,Q_0^2,a_0) &= a_s(Q^2,Q_0^2,a_0)\,.\cr
}\anoneqn$$
where $Q_1$ lies in the interval $[Q_0,Q]$;
since the evolution is logarithmic, we would presumably
choose $Q_1 = \sqrt{Q Q_0}$.
Formally, we want to take the limit $n\rightarrow\infty$, so
that
$$\asu(Q^2,Q_0^2,a_0) = \lim_{n\rightarrow\infty} a_s^{[n]}(Q^2,Q_0^2,a_0)\,.
\anoneqn$$
In practice, however, this is a {\it completely\/} pointless exercise, because
the form~(\use\NewAlphaS) is already within one part in $10^4$
of the `exact' answer for $Q>Q_0$, so long as $Q_0\geq 1.5$~GeV;
in the other direction, for $Q_0 = M_Z$,
the fractional error is less than $2\cdot 10^{-4}$ for $Q>4$~GeV.
(The `exact' answer refers to the numerical solution of the implicit
NLO equation (\use\Implicit).)
As $Q$ grows for fixed $Q_0$, the fractional error reaches a maximum for
some $Q_m^2$, and
then falls off; for $Q_0 = 2$~GeV, it is as noted less than 1 part in $10^4$,
around $Q_m = 27$~GeV.

With a formula for $a_s$ that reproduces a direct integration of
the $\beta$ function in hand, we can turn to the evolution equations
for the parton distributions.  Equation~(\use\AsEqn),
for the Mellin transform of the evolved parton distribution, is identical
to the original $x$-space equation.  In the usual Mellin space
approach, one performs the further truncations discussed in
section~\use\EvolutionSection,  effectively dropping terms
which are ${\cal O}(a_s^3)$, to obtain the
usual solution~(\use\EvolutionOperator,\use\SingletEvolutionOperator).
I should stress that from a physical
point of view, to next-to-leading order
the truncated equations are no less valid than the original,
and the solutions qualify no less as NLO solutions.  What is of interest
is the {\it discrepancy\/} between the two solutions: one or both
has higher-order corrections at least as large as half the difference
between them.  In order
to study this difference, we must solve the untruncated equation~(\use\AsEqn).
Introduce an evolution operator just as in
equation~(\use\EvolutionOperatorDef).  In the non-singlet case, we can
integrate the equation directly, obtaining
$$
E^z(a_s,a_0) = \exp\LB\int_{a_0}^{a_s} {P^z(a)\over\beta(a)}\, da\RB\,.
\eqn\ExactGeneralSolution$$
In the case at hand,
$$\eqalign{
\int_{a_0}^{a_s} &{P^z(a)\over\beta(a)} \,da =
-\int_{a_0}^{a_s} {P_0^z + a P_1^z\over a (\beta_0 + \beta_1 a)}\,da\cr
&= -{P_0^z\over\beta_0}\ln\L{a_s\over a_0}\R
  - {\Pb_1^z\over\beta_1}\,
    \ln\L{\beta_0+\beta_1 a_s\over\beta_0+\beta_1 a_0}\R\,,\cr
}\eqn\ExactNSintegral$$
so that
$$
E^z(a_s,a_0) = \L {a_s\over a_0}\R^{-P_0^z/\beta_0}\,
\L{\beta_0+\beta_1 a_s\over\beta_0+\beta_1 a_0}\R^{-\Pb_1^z/\beta_1}
                     \,.
\eqn\ExactEvolution$$
It is worth noting that this evolution operator is not much more expensive
to evaluate on a computer than the conventional one,
eqn.~(\use\EvolutionOperator).

Now, although we can solve the equation in closed form in the non-singlet
case, we will not be able to do so in the singlet case, so it is useful
to understand what alternative means we have of computing the operator
in eqn.~(\use\ExactEvolution).  Since it is an evolution operator, we
can write it in the form
$$
E^z(a_s,a_0) = \Pi_{j=1}^n E^z(a_j,a_{j-1})\,,
\eqn\EvolutionSequence$$
with $a_n = a_s$, where (for example),
$$
a_j = a_0 + (a_s-a_0) j/n\,.
\anoneqn$$
Equation~(\use\EvolutionSequence) is, of course, exact; but can we take
advantage of it to approximate $E^z(a_j,a_{j-1})$?  We would then hope
to recover the full evolution operator by taking the $n\rightarrow\infty$
limit,
$$
E^z(a_s,a_0) = \lim_{n\rightarrow\infty}\Pi_{j=1}^n E^z(a_j,a_{j-1})\,.
\anoneqn$$
In fact, we can, but
there is a subtlety concerning the expansion parameter.  Define
$\delta a = (a_s-a_0)/n$, the total evolution length for each step in
eqn.~(\use\EvolutionSequence), and expand $E^z(a_j,a_{j-1})$ in $\delta a$,
$$
E^z(a_j,a_{j-1}) = 1 + e_1(a_{j}) \delta a
                     + e_2(a_{j}) (\delta a)^2 + \cdots
\anoneqn$$
We then find
$$\eqalign{
E^z(a_s,a_0) &= 1 + \delta a \sum_{j=1}^n e_1(a_j)
  + (\delta a)^2 \sum_{j_1,j_2=1\atop j_1\neq j_2}^n e_1(a_{j_1}) e_1(a_{j_2})
  + \cdots +(\delta a)^n e_1(a_1)\cdots e_1(a_n)\cr
&\hskip 3mm
   + (\delta a)^2 \sum_{j=1}^n e_2(a_j)
  + (\delta a)^4 \sum_{j_1,j_2=1\atop j_1\neq j_2}^n e_2(a_{j_1}) e_2(a_{j_2})
  + \cdots +(\delta a)^{2n} e_2(a_1)\cdots e_2(a_n) + \cdots\cr
&= \L 1+\delta a\sum_{j=1}^n e_1(a_j)\R^n
   + {\rm terms\ with\ fewer\ sums\ than\ powers\ of\ }\delta a\,.
}\anoneqn$$
If the $e_i$ are well-behaved functions (as they are in our
case), roughly speaking each complete sum over an index produces
a factor of $n$ compensating the $1/n$ implicit in $\delta a$,
so that when we take the limit the terms with fewer sums than powers
will vanish because they are suppressed by powers of $n$.  Thus in the
limit we obtain
$$\eqalign{
E^z(a_s,a_0) &= \lim_{n\rightarrow\infty}
                    \L 1+\delta a\sum_{j=1}^n e_1(a_j)\R^n \cr
             &= \exp\LB \int_{a_0}^{a_s} da \; {d\ln E^z(a,a_0)\over da}\RB\cr
}\anoneqn$$
which using (\use\AsEqn)
is of course equivalent to eqn.~(\use\ExactGeneralSolution).
So we learn that expansion in $\delta a$ is legitimate, and that
the higher-order terms don't matter in the limit (they would of course
accelerate convergence if they were present).  This is neither
surprising nor subtle; the subtlety comes when we consider in addition
expanding in $a_j$, which is also a small parameter,
$$
e_i(a) = e_{i,0} + e_{i,1} a + e_{i,2} a^2 + \cdots
\anoneqn$$
and truncating at order $m$, to obtain $\tilde e_i(a)$.
Running through the above argument, we find that no matter how large
the number of segments $n$ gets, there is always an error of
${\cal O}(a^{m+1})$ in the estimate of $E^z(a_s,a_0)$,
$$\eqalign{
\tilde E^z(a_s,a_0) &= \lim_{n\rightarrow \infty}
                    \L 1+\delta a\sum_{j=1}^n \tilde e_1(a_j)\R^n \cr
     &= \exp\LB \int_{a_0}^{a_s} da \; {d\ln E^z(a,a_0)\over da}\RB
              +{\cal O}(a^{m+1})\cr
}\anoneqn$$
Unfortunately, the standard evolution
operator~(\use\EvolutionOperator)
{\it does\/} involve precisely these
sorts of truncations, so we can't use it to recover the missing
higher-order terms.  We can obtain a kernel that {\it will\/} work by
exponentiating only the leading-order term
of the integral~(\use\ExactNSintegral), and leaving the rest expanded,
$$\eqalign{
E^z(a_s,a_0) &=
\L {a_s\over a_0}\R^{-P_0^z/\beta_0}\,
\LB 1 -{\Pb_1^z\over\beta_1}\,
          \ln\L{\beta_0+\beta_1 a_s\over\beta_0+\beta_1 a_0}\R\RB \,.
}\eqn\ApproximateEvolution$$
This is equivalent to (formally) expanding in
$\Pb_1^z$, which is in general {\it not\/} a small
parameter; the expansion only works because the accompanying logarithm
is small.  We may thus expect that the speed of convergence varies as
we move around the complex plane, and
the question then arises of how large $n$ has to be in
equation~(\use\EvolutionSequence) before
we converge to the `exact' answer given by eqn.~(\use\ExactEvolution).
The answer, somewhat amusingly, is that for the non-singlet parton distribution
functions one encounters in practice,
a real-world value of $a_s$,
and evolution over the range from $Q_0 = 2 {\rm\ GeV}$ to $Q=15 {\rm\ GeV}$,
$n=1$ suffices (except at the smallest values of $x$, for which we need
$n=3$).  That is, using eqn.~(\use\ApproximateEvolution)
{\it already\/} gives the limiting answer, to within 2 parts in $10^4$.
More generally, $n=2$ limits the error to 6 parts in $10^4$ for $Q\leq 10 {\rm\
TeV}$.

In the singlet sector, equation~(\use\AsEqn) has the solution
$$
E^z(a_s,a_0) = T_a \exp\LB
\int_{a_0}^{a_s} {P^z(a)\over\beta(a)}\, da\RB\,.
\eqn\FormalSingletEvolution$$
where the $a$-ordering operator $T_a$, the analog of the usual path-ordering
operator, orders matrices $P^z(a)$ according to their distance
from $a_0$, putting
matrices of arguments further away from $a_0$ further to the left.  It appears
because in general $P^z$ matrices at different points $a,a'$ do not
commute.

Equation~(\use\FormalSingletEvolution) is still a formal expression, and one
must make further approximations to obtain an explicit expression.  We
can make use of eqn.~(\use\EvolutionSequence), in which
for singlet evolution we must also order the matrices,
$$
E^z(a_s,a_0) = E^z(a_n,a_{n-1})\cdots E^z(a_1,a_0)\,.
\eqn\SingletEvolutionSequence$$
Let us take a $\delta a$ sufficiently small that we can formally expand
NLO terms in the $a$-ordered
exponential~(\use\FormalSingletEvolution) to first order, without assuming
that the same is true of LO terms; we can then write
$$\eqalign{
E^z(a_s,a_0) &= T_a \left\{
\exp\LB -{P_0^z\over\beta_0} \int_{a_0}^{a_s} {da\over a}\RB\,
\exp\LB \int_{a_0}^{a_s} {P^z(a)\over\beta(a)}-\L{P_0^z\over -\beta_0 a}\R\,
                da\RB
 \right\}\,\cr
&= T_a \left\{
\exp\LB -{P_0^z\over\beta_0} \int_{a_0}^{a_s} {da\over a}\RB\,
   \L 1+ \int_{a_0}^{a_s} {P^z(a)\over\beta(a)}-\L{P_0^z\over -\beta_0 a}\R\,
                da\R
 \right\}\,\cr
}\eqn\TimeOrderedExpansion$$
Evaluating the time-ordered product gives us our final expression for
the singlet evolution operator,
$$\eqalign{
E^z&(a_s,a_0) = \cr
& \L{a_s\over a_0}\R^{-\lambda_{-}^z/\beta_0}\,\Bigg\{ P_{-}
 - {1\over\beta_1} P_{-} \Pb_1^z P_{-}\,
             \ln\LB{\beta_0+\beta_1 a_s\over\beta_0+\beta_1 a_0}\RB\cr
&\hskip 25mm
 -  P_{-} \Pb_1^z P_{+}
   \,{1\over \beta_0-\lambda_{+}^z+\lambda_{-}^z}\,\cr
&\hskip 17mm\times
      \LB a_s \L{a_s\over a_0}\R^{(\lambda_{-}^z-\lambda_{+}^z)/\beta_0}\!
           \F21\L 1,1-{\lambda_{+}^z-\lambda_{-}^z\over\beta_0};
               2-{\lambda_{+}^z-\lambda_{-}^z\over\beta_0};
             -{\beta_1\over\beta_0} a_s\R\RP\cr
&\hskip 35mm\LP -a_0 \,
           \F21\L 1,1-{\lambda_{+}^z-\lambda_{-}^z\over\beta_0};
               2-{\lambda_{+}^z-\lambda_{-}^z\over\beta_0};
             -{\beta_1\over\beta_0} a_0\R
        \vphantom{ \L{a_s\over a_0}\R^{(\lambda_{-}-\lambda_{+}^z)/\beta_0} }
            \RB\Bigg\}\cr
&\hskip 20mm + \L + \leftrightarrow -\R\cr
}\eqn\SingletApproximateEvolutionO$$
In this equation, $\F21$ is the hypergeometric function; while
$\beta_1$ is substantially larger than $\beta_0$, the argument
of the hypergeometric function will still be much smaller (in absolute
magnitude) than one, so that we can evaluate it using its series expansion,
$$
\F21\L a,b;c;z\R = {\Gamma(c)\over\Gamma(a)\Gamma(b)}\sum_{n=0}^\infty
  {\Gamma(n+a)\Gamma(n+b)\over\Gamma(n+c)\,n!} z^n
\anoneqn$$
To reproduce the toy parton as discussed in section~\use\NumericalPerformance,
we would need a rather large number of subintervals $n$
in eqn.~(\use\SingletEvolutionSequence)
for evolution at small and large $x$.  It is possible to reduce this
number by resumming the terms in $P_1$ that are proportional to $P_0$.
To do so, define
$$
\mu_{-}^z = \Tr\L  P_{-} \Pb_1^z \R\,,\hskip 1cm
\mu_{+}^z = \Tr\L  P_{+} \Pb_1^z \R\,.
\anoneqn$$
Then
$$ \Pb_1^z = \mu_{-}^z P_{-}+P_{-}\Pb_1^z P_{+}
+\mu_{+}^z P_{+}+P_{+}\Pb_1^z P_{-}\,,
\anoneqn$$
and
$$\eqalign{
E^z&(a_s,a_0) = \cr
& \L{a_s\over a_0}\R^{-\lambda_{-}^z/\beta_0}\,
  \L{\beta_0+\beta_1 a_s\over\beta_0+\beta_1 a_0}\R^{-\mu_{-}^z/\beta_1}\,
   \Bigg\{ P_{-}
 -  P_{-} \Pb_1^z P_{+}
   \,{1\over \beta_0-\lambda_{+}^z+\lambda_{-}^z}\,
     \L1+{\beta_1\over\beta_0} a_0\R^{(\mu_{+}^z-\mu_{-}^z)/\beta_1}\cr
&\hskip 17mm\times
      \LB a_s \L{a_s\over a_0}\R^{(\lambda_{-}^z-\lambda_{+}^z)/\beta_0}\!
           \F21\L 1+\mu_{+}^z-\mu_{-}^z,
               1-{\lambda_{+}^z-\lambda_{-}^z\over\beta_0};
               2-{\lambda_{+}^z-\lambda_{-}^z\over\beta_0};
             -{\beta_1\over\beta_0} a_s\R\RP\cr
&\hskip 35mm\LP -a_0 \,
           \F21\L 1+\mu_{+}^z-\mu_{-}^z,
               1-{\lambda_{+}^z-\lambda_{-}^z\over\beta_0};
               2-{\lambda_{+}^z-\lambda_{-}^z\over\beta_0};
             -{\beta_1\over\beta_0} a_0\R
        \vphantom{ \L{a_s\over a_0}\R^{(\lambda_{-}-\lambda_{+}^z)/\beta_0} }
            \RB\Bigg\}\cr
&\hskip 20mm + \L + \leftrightarrow -\R\cr
}\eqn\SingletApproximateEvolution$$
With this form, 20 subintervals suffice at $x\sim 10^{-5}$, 10 or fewer
at intermediate $x$, and only at the largest $x\sim0.7$ do we still
require a large number ($>100$) of subintervals.

Using equations~(\use\NewAlphaS,\use\ExactEvolution,%
\use\SingletApproximateEvolution,\use\SingletEvolutionSequence)
yields the evolution as would be given by a direct integration
of the untruncated evolution equation~(\use\OriginalEvolution)
along with a direct integration
of the beta function~(\use\BetaFunction).
One might have thought that this would be
the evolution as computed by various $x$-space programs; but it isn't.
The reason is that these programs do {\it not\/} appear to integrate
the beta function numerically, but rather use the approximate
solution~(\use\ApproximateAlphaS).  This provides yet a third inequivalent
version of NLO evolution.  It isn't difficult to reproduce it in the
Mellin approach, however; avoid changing variables to $a_s$, and
instead integrate eqn.~(\use\MellinFullEquation) with respect to $Q^2$.
In the non-singlet sector, we obtain
$$\eqalign{
\hat E^z&(Q^2,Q_0^2) = \cr
&\L{L_0\over L}\R^{-P_0^z/\beta_0}\,
\exp\LB {1\over\beta_0^2} \Pb_1^z \,\L {1\over L_0}-{1\over L}\R
     -{\beta_1\over\beta_0^3} P_0^z \,\L {\ln L_0\over L_0}-{\ln L\over L}\R
\RP\cr&\hskip 35mm\LP
     -{\beta_1\over2\beta_0^4} P_1^z \,\L {1+2\ln L_0\over L_0^2}
                                         -{1+2\ln L\over L^2}\R
\RP\cr&\hskip 35mm\LP
     +{\beta_1^2\over27\beta_0^6} P_1^z \,\L {2+6\ln L_0+9\ln^2 L_0\over L_0^3}
                                         -{2+6\ln L+9\ln^2 L\over L^3}\R\RB
}\anoneqn$$
where the hat on $E$ indicates that its arguments are momentum scales
instead of running couplings, and where $L_i = \ln Q_i^2/\Lambda^2$.
We can obtain the analog to equation~(\use\ApproximateEvolution) by
formally expanding the exponential in the $P_i^z$, but leaving the power
prefactor unexpanded.  In the singlet sector, the
first-order expansion in eqn.~(\use\TimeOrderedExpansion) yields
\def\rL{r_L^{\delta^z}}
\def\nl{\cr &\hskip 10mm}
$$\hskip -15pt\eqalign{
&\hat E^z(Q^2,Q_0^2) =
r_L^{-\lambda_{-}^z/\beta_0}\cr
&\hskip 3mm\times\Bigg\{ P_{-}
 + P_{-} \LB {1\over\beta_0^2} \Pb_1^z \,\L {1\over L_0}-{1\over L}\R
   -{\beta_1\over\beta_0^3} P_0^z \,\L {\ln L_0\over L_0}-{\ln L\over L}\R
     -{\beta_1\over2\beta_0^4} P_1^z \,\L {1+2\ln L_0\over L_0^2}
                                         -{1+2\ln L\over L^2}\R
\RP\cr&\hskip 27mm\LP
     +{\beta_1^2\over27\beta_0^6} P_1^z \,\L {2+6\ln L_0+9\ln^2 L_0\over L_0^3}
                               -{2+6\ln L+9\ln^2 L\over L^3}\R\RB P_{-}\,\cr
&\hskip 4mm
 +  P_{-} \LB
   {1\over\beta_0 (\beta_0+\lambda_{-}^z\!-\lambda_{+}^z)}
       P_1^z
            \L {1\over L_0}-{1\over L} \rL\R\RP
   -2 {\beta_1\over \beta_0^3 (2 \beta_0+\lambda_{-}^z\!-\lambda_{+}^z)}
         P_1^z \L {\ln L_0\over  L_0^2} -{\ln L\over L^2} \rL\R
 \nl
   -2 {\beta_1\over\beta_0^2 (2 \beta_0+\lambda_{-}^z-\lambda_{+}^z)^2}
         P_1^z \L {1\over L_0^2} -{1\over L^2} \rL\R
   +{\beta_1^2\over \beta_0^5 (3 \beta_0+\lambda_{-}^z-\lambda_{+}^z)}
         P_1^z \L {\ln^2 L_0\over L_0^3} -{\ln^2 L\over L^3} \rL\R
 \nl
   +2 {\beta_1^2\over\beta_0^4 (3 \beta_0+\lambda_{-}^z-\lambda_{+}^z)^2}
         P_1^z \L{\ln L_0\over L_0^3} -{\ln L\over L^3} \rL\R
  \LP
   +2 {\beta_1^2\over \beta_0^3 (3 \beta_0+\lambda_{-}^z-\lambda_{+}^z)^3}
         P_1^z \L {1\over L_0^3} -{1\over L^3} \rL\R\RB
       \,P_{+} \Bigg\}\cr
&\hskip 20mm + \L + \leftrightarrow -\R
        \vphantom{ \L{a_s\over a_0}\R^{(\lambda_{-}-\lambda_{+}^z)/\beta_0} }
\cr
}\anoneqn$$
where $r_L = L/L_0$ and
$\delta^z \equiv (\lambda_{+}^z-\lambda_{-}^z)/\beta_0$.
This form of the evolution operator should again be used in combination
with eqn.~(\use\SingletEvolutionSequence); it does require
a rather large number of subintervals $n$ if we want
an answer accurate to 2 parts in 10${}^4$.  It is
thus better to add in the terms arising from an expansion of the time-ordered
exponential to second order,
$$\eqalign{
&\hat \delta E_2^z(Q^2,Q_0^2) = \sum_{\sigma_{1,2,3}=\pm}
   P_{\sigma_1}^z  T_2^z(\lambda_{\sigma_1}^z,\lambda_{\sigma_2}^z,
                      \lambda_{\sigma_3}^z,P_{\sigma_2}^z)  P_{\sigma_3}^z\,.
}\eqn\SecondOrderSingletEvolution$$
The (somewhat lengthy) formula for $T_2$ is given in
appendix~\use\SecondOrderTerms; it should be kept in mind that most of
the computation need only be done once, not anew for each value of $Q^2$.

With the addition of the
terms in $T_2$, most values of $x$ require $n=3$ or $4$
in eqn.~(\use\SingletEvolutionSequence), though at $x=0.7$, 30
subintervals are required to provide an accuracy of $2\cdot 10^{-4}$
for the toy parton set of ref.~[\use\Compare].

Using these evolution operators, along with eqn.~(\use\ApproximateAlphaS), I
find that I indeed reproduce the results of the `direct solution,' that
is the upper half of table~1 in ref.~[\use\Compare], mostly within their
quoted errors.  Curiously enough,
the values of the evolved parton densities are quite close to what
would be obtained from the `exact' evolution described earlier in
this section, typically within a few parts in $10^3$.
Note, however, that the running coupling values are
somewhat different: with $Q_0 = 2$~GeV, the two values for the running
coupling differ by about 1\% at $Q=10$~GeV.

\section{Conclusions}
\tagsection\Conclusions
\vskip 10pt

In this paper, I have derived a quadratic contour for calculating the evolution
of parton distribution functions within the Mellin transform method,
and demonstrated its superiority over other techniques in the literature.  I
have
also shown how to reproduce the results obtained within the `$x$-space' method
using a modified evolution operator.  In addition to the application discussed
here, the method may also be used within a general framework for extracting
parton distribution functions from collider data~[\use\Extraction].

I thank M. Peskin and A. Vogt for helpful comments.

\appendix{Analytic Continuation of a Dilogarithm Integral}
\tagappendix\DilogMomentAppendix
\vskip 10pt

\def\Li{{\rm Li}}
In order to evaluate the evolution operators along our chosen contours
in the complex plane, we must be able to evaluate the moments of the
leading- and next-to-leading
order anomalous dimensions at essentially arbitrary points in the complex
plane.
As discussed in ref.~[\GRV], all functions appearing in the Mellin
moments of these anomalous dimensions, save one,
have expressions in terms of elementary
functions or derivatives of the $\Gamma$ function.  The latter can
be calculated everywhere in the complex plane via well-known techniques.
The one exception is ${\tilde S}(z)$, for which GRV give the following
expression,
$$
{\tilde S}(z) = -{5\over 8}\zeta(3) + \eta
  \,\LB {(\psi(z+1)-\psi(1))\over z^2}
        - {\zeta(2)\over2} \L\psi((z+1)/2)-\psi(z/2)\R
        +\int_0^1 dx\,x^{z-1} {\Li_2(x)\over 1+x}\RB
\anoneqn$$
where $\eta = \pm 1$, $\psi$ is the usual logarithmic derivative of
the gamma function,
and $\Li_2$ is the dilogarithm.  As it is not completely clear that
the expansion given by the authors of ref.~[\use\GRV] for the last term
is valid for our purposes, I give an alternate one here.

Define
$$M_1(z+1) = \int_0^1 dx\,x^{z-1} {\Li_2(x)\over 1+x}\,.
\anoneqn$$
Using $x^{z-1} = (1+x) x^{z-2} - x^{z-2}$, we find the following
recurrence relation for $M_1$,
$$\eqalign{
M_1(z+1) &= -M_1(z) + \int_0^1 dx\, x^{z-2} \Li_2(x)\cr
  &= M_1(z-1) -\int_0^1 dx\, x^{z-3} (1-x)\Li_2(x)\,.
}\eqn\MoneRecurrence$$
The other integrals in this equation may be performed by integration by parts.
Let
$$\eqalign{
M_2(z) &\equiv \int_0^1 dx\, x^{z-2} \Li_2(x)
= {\zeta(2)\over z-1} + {1\over z-1}\int_0^1 dx\,x^{z-2} \ln(1-x)\cr
&={\zeta(2)\over z-1} + {1\over z-1}\LP{d\over da}\RV_{a=0}
   \int_0^1 dx\,x^{z-2} (1-x)^a\cr
&={\zeta(2)\over z-1} + {1\over (z-1)^2}\L \psi(1)-\psi(z)\R\,;\cr
}\anoneqn$$
then
$$\eqalign{
M_2(z)-M_2(z-1) &=
-\int_0^1 dx\, x^{z-3} (1-x)\Li_2(x)\cr
&= -{\zeta(2)\over (z-1)(z-2)}
  - \L{1\over (z-2)^2}-{1\over (z-1)^2}\R\,\L\psi(z-1)-\psi(1)\R
  + {1\over (z-1)^3}\,.
}\anoneqn$$

On the other hand, we can also write down an `asymptotic' expansion for
$M_1$; to do so, repeatedly rewrite
$$
{1\over 1+x} = {1-x\over 2(1+x)} + {1\over2}\,,
\anoneqn$$
to obtain
$$\eqalign{
M_1(z) &= {1\over2}\int_0^1 dx\,x^{z-2} \Li_2(x)
        + {1\over2}\int_0^1 dx\,x^{z-2} (1-x) {\Li_2(x)\over 1+x}\cr
     &= {1\over2}\sum_{j=0}^N 2^{-j}\int_0^1 dx\,x^{z-2} (1-x)^j\Li_2(x)
        + 2^{-N-1}\int_0^1 dx\,x^{z-2} (1-x)^{N+1} {\Li_2(x)\over 1+x}\cr
     &= {1\over2}\sum_{j=0}^N 2^{-j}\sum_{l=0}^j (-1)^l \L{j\atop l}\R M_2(z+l)
        + 2^{-N-1}\int_0^1 dx\,x^{z-2} (1-x)^{N+1} {\Li_2(x)\over 1+x}\cr
}\eqn\ExpandMone$$
Since in the $z\rightarrow\infty$ limit, the integral is dominated
by the region $x\sim 1$, the last term goes as $1/z^{N+2}$, and we can
drop it if we are interested in the asymptotic expansion only through
order $N+1$.  In fact, because of the factor of $2^{-N}$ in front,
dropping this term is a reasonable approximation even for modest $z$.
We can re-expand the resulting expression in terms of $1/z$, but
this yields a rather poor representation.  Instead we can rewrite
the remainder term
$$\eqalign{
 2^{-N-1}&\int_0^1 dx\,x^{z-2} (1-x)^{N+1} {\Li_2(x)\over 1+x}
= \cr
 &2^{-N-2}\Li_2(1)\int_0^1 dx\,x^{z-1} (1-x)^{N+1}
+2^{-N-1}\int_0^1 dx\,x^{z-2} (1-x)^{N+1}
 \L{\Li_2(x)\over 1+x}-{x \Li_2(1)\over2}\R\cr
&= 2^{-N-2}\zeta(2){\Gamma(z)\Gamma(N+2)\over\Gamma(z+N+2)}
+2^{-N-1}\int_0^1 dx\,x^{z-2} (1-x)^{N+1}
 \L{\Li_2(x)\over 1+x}-{x \Li_2(1)\over2}\R\cr
}\eqn\IntegralFit$$
The second integral goes as $\ln z/z^{N+3}$.  For $\Re z$ sufficiently
large, we simply drop the second term; for other $z$, we can use the
recurrence relation~(\use\MoneRecurrence) [backwards]
to shift $z$ into this range.
It's worth using the recurrence
relation~(\use\MoneRecurrence) a few times explicitly, since we will
certainly be interested in values near $z\sim 1$~or~$2$ for which
the approximation with reasonable $N$ won't be enough; we can then simplify
the transcendental functions to minimize the number of such function
evaluations we need to perform.

\appendix{Second-Order Terms for the $x$-Space Singlet Evolution Operator}
\tagappendix\SecondOrderTerms
\vskip 10pt

To present the (somewhat lengthy) formula for $T_2$, used
in equation~(\use\SecondOrderSingletEvolution), define the
following functions,
$$\eqalign{
d_{c;ij} &= c + {\lambda_i^z-\lambda_j^z\over\beta_0}\,,\cr
P_{11} &= {1\over\beta_0^2} P_1 P P_1\,,\cr
P_{10} &= {\beta_1\over\beta_0^3} P_1 P P_0\,,\cr
P_{01} &= {\beta_1\over\beta_0^3} P_0 P P_1\,,\cr
P_{00} &= {\beta_1^2\over\beta_0^4} P_0 P P_0\,,\cr
P_{11}^{(a)} &= P_{11}- d_{1;23} P_{10}\,,\cr
P_{11}^{(b)} &= P_{11}- d_{1;12} P_{01}\,,\cr
P_{01}^{(a)} &= P_{01}- d_{1;23} P_{00}\,,\cr
P_{10}^{(a)} &= P_{10}- d_{1;12} P_{00}\,,\cr
\delta_i^z &= \lambda_i^z/\beta_0\,.\cr
}\anoneqn$$
As in section~\use\XSpaceSection, $r_L = L/L_0$.

Then
\def\la#1{\lambda_{#1}}\def\d#1{\delta_{#1}^z}

\def\split{\vphantom{P_{11}^{(a)}}\RP\cr&\hskip 30mm\LP\vphantom{P_{11}^{(a)}}}
$$\hskip -20pt\eqalign{
&T_2^z(\la1,\la2,\la3,P) =
-{\frac{d_{1;23}\ln {L_0} {r}_{L}^{\d1}}{\beta_{0}^{2}L_{0}^{2}}}
   \left( {P}_{1}^{(a)}{{{d}}_{1;12}} + {P}_{10}^{(a)}{{{d}}_{1;12}} -
{{{P}}_{10}}{d_{2;13}} - {P}_{1}^{(a)}{{{d}}_{2;13}} + 2{P_{00}}d_{2;13}^{2}
\right)
\cr &\hskip 2mm
+ {\frac{{d_{1;23}} r_{L}^{\d1}}{\beta_{0}^{2}L_{0}^{2}}}
   \left( P_{11}^{(a)}{d_{1;12}} - P_{11}^{(a)}{d_{2;13}} -
P_{01}^{(a)}d_{1;12}^{2} + {P_{10}}d_{2;13}^{2} + P_{01}^{(a)}d_{2;13}^{2} -
2{P_{00}}d_{2;13}^{3} \right)
\cr &\hskip 2mm
- {\frac{2{{\beta}_1}\ln {L_0} {r}_{L}^{\d1}}{\beta_{0}^{4}L_{0}^{3}}}
   \left( {P}_{11}^{(a)}{{{d}}_{1;23}}{d_{2;12}} +
{P}_{11}^{(b)}{{{d}}_{1;12}}{d_{2;23}} - {P}_{11}^{(a)}{{{d}}_{1;23}}{d_{3;13}}
- {{{P}}_{11}}{d_{2;23}}{d_{3;13}} - {{{P}}_{10}}{d_{1;23}}d_{2;12}^{2}
  \split
   - {{{P}}_{01}}{d_{1;12}}d_{2;23}^{2} + {{{P}}_{01}}{d_{3;13}}d_{2;23}^{2} +
2{P_{10}}{d_{1;23}}d_{3;13}^{2} + 2{P_{01}}{d_{2;23}}d_{3;13}^{2} \right)
\cr &\hskip 2mm
- {\frac{2{{\beta}_1} r_{L}^{\d1}}{\beta_{0}^{4}L_{0}^{3}}}
   \left( P_{11}^{(a)}{d_{1;23}}d_{2;12}^{2} +
P_{11}^{(b)}{d_{1;12}}d_{2;23}^{2} - {P_{11}}{d_{3;13}}d_{2;23}^{2} -
P_{11}^{(a)}{d_{1;23}}d_{3;13}^{2} - {P_{11}}{d_{2;23}}d_{3;13}^{2}
  \split
  + {P_{01}}d_{2;23}^{2}d_{3;13}^{2}
+ 2{P_{10}}{d_{1;23}}d_{3;13}^{3} + 2{P_{01}}{d_{2;23}}d_{3;13}^{3} \right)
\cr &\hskip 2mm
+ {\frac{2\ln {L_0} \beta_{1}^{2}{r}_{L}^{\d1}}{\beta_{0}^{6}L_{0}^{4}}}
   \left( 2{P_{11}}{d_{2;23}}d_{2;12}^{2} + 2{P_{11}}{d_{2;12}}d_{2;23}^{2} -
2{P_{11}}{d_{4;13}}d_{2;23}^{2} + {P}_{11}^{(a)}{{{d}}_{1;23}}d_{3;12}^{2} -
{{{P}}_{10}}{d_{1;23}}d_{3;12}^{3}
  \split\hskip -5mm
+ {P}_{11}^{(b)}{{{d}}_{1;12}}d_{3;23}^{2} - {{{P}}_{11}}{d_{4;13}}d_{3;23}^{2}
- {{{P}}_{01}}{d_{1;12}}d_{3;23}^{3} + {{{P}}_{01}}{d_{4;13}}d_{3;23}^{3} -
{P}_{11}^{(a)}{{{d}}_{1;23}}d_{4;13}^{2}
  \split\hskip -5mm
 - 4{P_{11}}{d_{2;23}}d_{4;13}^{2} - {{{P}}_{11}}{d_{3;23}}d_{4;13}^{2} +
2{P_{01}}d_{3;23}^{2}d_{4;13}^{2} + 3{P_{10}}{d_{1;23}}d_{4;13}^{3} +
3{P_{01}}{d_{3;23}}d_{4;13}^{3} \right)
\cr &\hskip 2mm
+ {\frac{2 \beta_{1}^{2}r_{L}^{\d1}}{\beta_{0}^{6}L_{0}^{4}}}
   \left( 2{P_{11}}d_{2;12}^{2}d_{2;23}^{2} +
P_{11}^{(a)}{d_{1;23}}d_{3;12}^{3} + P_{11}^{(b)}{d_{1;12}}d_{3;23}^{3} -
{P_{11}}{d_{4;13}}d_{3;23}^{3} - 2{P_{11}}d_{2;23}^{2}d_{4;13}^{2}
  \split
- {P_{11}}d_{3;23}^{2}d_{4;13}^{2} + {P_{01}}d_{3;23}^{3}d_{4;13}^{2} -
P_{11}^{(a)}{d_{1;23}}d_{4;13}^{3} - 4{P_{11}}{d_{2;23}}d_{4;13}^{3}
  \split
 - {P_{11}}{d_{3;23}}d_{4;13}^{3}
+ 2{P_{01}}d_{3;23}^{2}d_{4;13}^{3} + 3{P_{10}}{d_{1;23}}d_{4;13}^{4} +
3{P_{01}}{d_{3;23}}d_{4;13}^{4} \right)
\cr &\hskip 2mm
- {\frac{4{P_{11}}\ln {L_0}
\beta_{1}^{3}{r}_{L}^{\d1}}{\beta_{0}^{8}L_{0}^{5}}}
   \left( {d}_{2;23}^{2}{d}_{3;12}^{2} + {{{d}}_{2;23}}d_{3;12}^{3} +
{d}_{2;12}^{2}{d}_{3;23}^{2} + {{{d}}_{2;12}}d_{3;23}^{3} -
{{{d}}_{5;13}}d_{3;23}^{3} - {d}_{2;23}^{2}{d}_{5;13}^{2}
  \split
- 2d_{3;23}^{2}d_{5;13}^{2}
- 3{d_{2;23}}d_{5;13}^{3} - 3{d_{3;23}}d_{5;13}^{3} \right)
+ {\frac{{P_{11}}{d_{3;23}}{d_{6;13}}\beta_{1}^{4}r_{L}^{{\d3}}{{\ln
}^4}{L}}{{L^6}\beta_{0}^{10}}}
\cr &\hskip 2mm
- {\frac{4{P_{11}}\beta_{1}^{3}r_{L}^{\d1}}{\beta_{0}^{8}L_{0}^{5}}}
   \left( d_{2;23}^{2}d_{3;12}^{3} + d_{2;12}^{2}d_{3;23}^{3} -
d_{3;23}^{3}d_{5;13}^{2} - d_{2;23}^{2}d_{5;13}^{3} - 2d_{3;23}^{2}d_{5;13}^{3}
- 3{d_{2;23}}d_{5;13}^{4} - 3{d_{3;23}}d_{5;13}^{4} \right)
\cr &\hskip 2mm
+ {\frac{4{P_{11}}{d_{3;23}}\ln {L_0}
\beta_{1}^{4}{r}_{L}^{\d1}}{\beta_{0}^{10}L_{0}^{6}}}
   \left( {{{d}}_{3;23}}d_{3;12}^{3} + {d}_{3;12}^{2}{d}_{3;23}^{2} -
{d}_{3;23}^{2}{d}_{6;13}^{2} - 3{d_{3;23}}d_{6;13}^{3} - 6d_{6;13}^{4} \right)
\cr &\hskip 2mm
+ {\frac{4{P_{11}}{d_{3;23}}
\beta_{1}^{4}r_{L}^{\d1}}{\beta_{0}^{10}L_{0}^{6}}}
   \left( d_{3;12}^{3}d_{3;23}^{2} - d_{3;23}^{2}d_{6;13}^{3} -
3{d_{3;23}}d_{6;13}^{4} - 6d_{6;13}^{5} \right)
\cr &\hskip 2mm
- {\frac{{d_{1;12}} {d_{1;23}}r_{L}^{{\d2}}}{L\beta_{0}^{2}L_{0}}}
   \left( P_{11}^{(a)} - P_{01}^{(a)}{d_{1;12}} \right)
+ {\frac{P_{01}^{(a)}{d_{1;12}}{d_{1;23}}\ln
{L}{r}_{L}^{{\d2}}}{L\beta_{0}^{2}L_{0}}}
+ {\frac{P_{10}^{(a)}{d_{1;12}}{d_{1;23}}\ln
{L_0}{r}_{L}^{{\d2}}}{L\beta_{0}^{2}L_{0}}}
\cr &\hskip 2mm
- {\frac{{P_{00}}{d_{1;12}}{d_{1;23}}\ln {L}\ln
{L_0}{r}_{L}^{{\d2}}}{L\beta_{0}^{2}L_{0}}}
+ {\frac{2{{\beta}_1}P_{11}^{(a)}{d_{1;23}}{d_{2;12}}\ln
{L}{r}_{L}^{{\d2}}}{{L^2}\beta_{0}^{4}L_{0}}}
- {\frac{2{{\beta}_1}{P_{10}}{d_{1;23}}{d_{2;12}}\ln {L}\ln
{L_0}{r}_{L}^{{\d2}}}{{L^2}\beta_{0}^{4}L_{0}}}
\cr &\hskip 2mm
+
{\frac{2{{\beta}_1}P_{11}^{(a)}{d_{1;23}}d_{2;12}^{2}r_{L}^{{\d2}}}{{L^2}\beta_{0}^{4}L_{0}}}
- {\frac{2{{\beta}_1}{P_{10}}{d_{1;23}}\ln
{L_0}{d}_{2;12}^{2}{r}_{L}^{{\d2}}}{{L^2}\beta_{0}^{4}L_{0}}}
- {\frac{2P_{11}^{(a)}{d_{1;23}}\ln
{L}{d}_{3;12}^{2}\beta_{1}^{2}{r}_{L}^{{\d2}}}{{L^3}\beta_{0}^{6}L_{0}}}
\cr &\hskip 2mm
+ {\frac{2{P_{10}}{d_{1;23}}\ln {L}\ln
{L_0}{d}_{3;12}^{2}\beta_{1}^{2}{r}_{L}^{{\d2}}}{{L^3}\beta_{0}^{6}L_{0}}}
-
{\frac{2P_{11}^{(a)}{d_{1;23}}d_{3;12}^{3}\beta_{1}^{2}r_{L}^{{\d2}}}{{L^3}\beta_{0}^{6}L_{0}}}
+ {\frac{2{P_{10}}{d_{1;23}}\ln
{L_0}{d}_{3;12}^{3}\beta_{1}^{2}{r}_{L}^{{\d2}}}{{L^3}\beta_{0}^{6}L_{0}}}
\cr
}$$

$$\hskip-20pt\eqalign{ &\hskip 2mm
+ {\frac{2{{\beta}_1}P_{11}^{(b)}{d_{1;12}}{d_{2;23}}\ln
{L_0}{r}_{L}^{{\d2}}}{L\beta_{0}^{4}L_{0}^{2}}}
- {\frac{2{{\beta}_1}{P_{01}}{d_{1;12}}{d_{2;23}}\ln {L}\ln
{L_0}{r}_{L}^{{\d2}}}{L\beta_{0}^{4}L_{0}^{2}}}
+
{\frac{2{{\beta}_1}P_{11}^{(b)}{d_{1;12}}d_{2;23}^{2}r_{L}^{{\d2}}}{L\beta_{0}^{4}L_{0}^{2}}}
\cr &\hskip 2mm
- {\frac{2{{\beta}_1}{P_{01}}{d_{1;12}}\ln
{L}{d}_{2;23}^{2}{r}_{L}^{{\d2}}}{L\beta_{0}^{4}L_{0}^{2}}}
- {\frac{4{P_{11}}{d_{2;12}}{d_{2;23}}\ln {L}\ln
{L_0}\beta_{1}^{2}{r}_{L}^{{\d2}}}{{L^2}\beta_{0}^{6}L_{0}^{2}}}
- {\frac{4{P_{11}}{d_{2;23}}\ln
{L_0}{d}_{2;12}^{2}\beta_{1}^{2}{r}_{L}^{{\d2}}}{{L^2}\beta_{0}^{6}L_{0}^{2}}}
\cr &\hskip 2mm
- {\frac{4{P_{11}}{d_{2;12}}\ln
{L}{d}_{2;23}^{2}\beta_{1}^{2}{r}_{L}^{{\d2}}}{{L^2}\beta_{0}^{6}L_{0}^{2}}}
-
{\frac{4{P_{11}}d_{2;12}^{2}d_{2;23}^{2}\beta_{1}^{2}r_{L}^{{\d2}}}{{L^2}\beta_{0}^{6}L_{0}^{2}}}
+ {\frac{4{P_{11}}{d_{2;23}}\ln {L}\ln
{L_0}{d}_{3;12}^{2}\beta_{1}^{3}{r}_{L}^{{\d2}}}{{L^3}\beta_{0}^{8}L_{0}^{2}}}
\cr &\hskip 2mm
+ {\frac{4{P_{11}}\ln
{L}{d}_{2;23}^{2}{d}_{3;12}^{2}\beta_{1}^{3}{r}_{L}^{{\d2}}}{{L^3}\beta_{0}^{8}L_{0}^{2}}}
+ {\frac{4{P_{11}}{d_{2;23}}\ln
{L_0}{d}_{3;12}^{3}\beta_{1}^{3}{r}_{L}^{{\d2}}}{{L^3}\beta_{0}^{8}L_{0}^{2}}}
+
{\frac{4{P_{11}}d_{2;23}^{2}d_{3;12}^{3}\beta_{1}^{3}r_{L}^{{\d2}}}{{L^3}\beta_{0}^{8}L_{0}^{2}}}
\cr &\hskip 2mm
- {\frac{2P_{11}^{(b)}{d_{1;12}}\ln
{L_0}{d}_{3;23}^{2}\beta_{1}^{2}{r}_{L}^{{\d2}}}{L\beta_{0}^{6}L_{0}^{3}}}
+ {\frac{2{P_{01}}{d_{1;12}}\ln {L}\ln
{L_0}{d}_{3;23}^{2}\beta_{1}^{2}{r}_{L}^{{\d2}}}{L\beta_{0}^{6}L_{0}^{3}}}
-
{\frac{2P_{11}^{(b)}{d_{1;12}}d_{3;23}^{3}\beta_{1}^{2}r_{L}^{{\d2}}}{L\beta_{0}^{6}L_{0}^{3}}}
\cr &\hskip 2mm
+ {\frac{2{P_{01}}{d_{1;12}}\ln
{L}{d}_{3;23}^{3}\beta_{1}^{2}{r}_{L}^{{\d2}}}{L\beta_{0}^{6}L_{0}^{3}}}
+ {\frac{4{P_{11}}{d_{2;12}}\ln {L}\ln
{L_0}{d}_{3;23}^{2}\beta_{1}^{3}{r}_{L}^{{\d2}}}{{L^2}\beta_{0}^{8}L_{0}^{3}}}
+ {\frac{4{P_{11}}\ln
{L_0}{d}_{2;12}^{2}{d}_{3;23}^{2}\beta_{1}^{3}{r}_{L}^{{\d2}}}{{L^2}\beta_{0}^{8}L_{0}^{3}}}
\cr &\hskip 2mm
+ {\frac{4{P_{11}}{d_{2;12}}\ln
{L}{d}_{3;23}^{3}\beta_{1}^{3}{r}_{L}^{{\d2}}}{{L^2}\beta_{0}^{8}L_{0}^{3}}}
+
{\frac{4{P_{11}}d_{2;12}^{2}d_{3;23}^{3}\beta_{1}^{3}r_{L}^{{\d2}}}{{L^2}\beta_{0}^{8}L_{0}^{3}}}
- {\frac{4{P_{11}}\ln {L}\ln
{L_0}{d}_{3;12}^{2}{d}_{3;23}^{2}\beta_{1}^{4}{r}_{L}^{{\d2}}}{{L^3}\beta_{0}^{10}L_{0}^{3}}}
\cr &\hskip 2mm
- {\frac{4{P_{11}}\ln
{L_0}{d}_{3;12}^{3}{d}_{3;23}^{2}\beta_{1}^{4}{r}_{L}^{{\d2}}}{{L^3}\beta_{0}^{10}L_{0}^{3}}}
- {\frac{4{P_{11}}\ln
{L}{d}_{3;12}^{2}{d}_{3;23}^{3}\beta_{1}^{4}{r}_{L}^{{\d2}}}{{L^3}\beta_{0}^{10}L_{0}^{3}}}
- {\frac{2{P_{11}}{d_{3;23}}\ln
{L}{d}_{3;12}^{2}\beta_{1}^{4}{r}_{L}^{{\d2}}{{\ln
}^2}{L_0}}{{L^3}\beta_{0}^{10}L_{0}^{3}}}
\cr &\hskip 2mm
- {\frac{{d_{1;23}}{d_{2;13}} \ln {L}{r}_{L}^{{\d3}}}{{L^2}\beta_{0}^{2}}}
   \left( {P_{10}} + P_{01}^{(a)} - 2{P_{00}}{d_{2;13}} \right)
+ {\frac{{d_{1;23}}{d_{2;13}} r_{L}^{{\d3}}}{{L^2}\beta_{0}^{2}}}
   \left( P_{11}^{(a)} - {P_{10}}{d_{2;13}} - P_{01}^{(a)}{d_{2;13}} +
2{P_{00}}d_{2;13}^{2} \right)
\cr &\hskip 2mm
- {\frac{2{{\beta}_1}{d_{3;13}}\ln {L} {r}_{L}^{{\d3}}}{{L^3}\beta_{0}^{4}}}
   \left( {P}_{11}^{(a)}{{{d}}_{1;23}} + {{{P}}_{11}}{d_{2;23}} -
2{P_{10}}{d_{1;23}}{d_{3;13}} - 2{P_{01}}{d_{2;23}}{d_{3;13}} -
{{{P}}_{01}}d_{2;23}^{2} \right)
\cr &\hskip 2mm
- {\frac{2{{\beta}_1}{d_{3;13}} r_{L}^{{\d3}}}{{L^3}\beta_{0}^{4}}}
   \left( P_{11}^{(a)}{d_{1;23}}{d_{3;13}} + {P_{11}}{d_{2;23}}{d_{3;13}} +
{P_{11}}d_{2;23}^{2} - {P_{01}}{d_{3;13}}d_{2;23}^{2} -
2{P_{10}}{d_{1;23}}d_{3;13}^{2} - 2{P_{01}}{d_{2;23}}d_{3;13}^{2} \right)
\cr &\hskip 2mm
- {\frac{P_{11}^{(a)}{d_{1;23}}{d_{3;12}}\beta_{1}^{2}r_{L}^{{\d2}}{{\ln
}^2}{L}}{{L^3}\beta_{0}^{6}L_{0}}}
+ {\frac{{P_{10}}{d_{1;23}}{d_{3;12}}\ln {L_0}\beta_{1}^{2}{r}_{L}^{{\d2}}{{\ln
}^2}{L}}{{L^3}\beta_{0}^{6}L_{0}}}
+ {\frac{2{P_{11}}{d_{2;23}}{d_{3;12}}\ln
{L_0}\beta_{1}^{3}{r}_{L}^{{\d2}}{{\ln }^2}{L}}{{L^3}\beta_{0}^{8}L_{0}^{2}}}
\cr &\hskip 2mm
+ {\frac{2{P_{11}}{d_{3;12}}d_{2;23}^{2}\beta_{1}^{3}r_{L}^{{\d2}}{{\ln
}^2}{L}}{{L^3}\beta_{0}^{8}L_{0}^{2}}}
- {\frac{2{P_{11}}{d_{3;12}}\ln
{L_0}{d}_{3;23}^{2}\beta_{1}^{4}{r}_{L}^{{\d2}}{{\ln
}^2}{L}}{{L^3}\beta_{0}^{10}L_{0}^{3}}}
- {\frac{2{P_{11}}{d_{3;12}}d_{3;23}^{3}\beta_{1}^{4}r_{L}^{{\d2}}{{\ln
}^2}{L}}{{L^3}\beta_{0}^{10}L_{0}^{3}}}
\cr &\hskip 2mm
- {\frac{P_{11}^{(b)}{d_{1;12}}{d_{3;23}}\beta_{1}^{2}r_{L}^{{\d2}}{{\ln
}^2}{L_0}}{L\beta_{0}^{6}L_{0}^{3}}}
+ {\frac{{P_{01}}{d_{1;12}}{d_{3;23}}\ln {L}\beta_{1}^{2}{r}_{L}^{{\d2}}{{\ln
}^2}{L_0}}{L\beta_{0}^{6}L_{0}^{3}}}
+ {\frac{2{P_{11}}{d_{2;12}}{d_{3;23}}\ln {L}\beta_{1}^{3}{r}_{L}^{{\d2}}{{\ln
}^2}{L_0}}{{L^2}\beta_{0}^{8}L_{0}^{3}}}
\cr &\hskip 2mm
- {\frac{{P_{11}}{d_{3;12}}{d_{3;23}}\beta_{1}^{4}r_{L}^{{\d2}}{{\ln
}^2}{L}{{\ln }^2}{L_0}}{{L^3}\beta_{0}^{10}L_{0}^{3}}}
- {\frac{ {d_{4;13}}\beta_{1}^{2}r_{L}^{{\d3}}{{\ln
}^3}{L}}{{L^4}\beta_{0}^{6}}}
   \left( {P_{10}}{d_{1;23}} + {P_{01}}{d_{3;23}} \right)
+ {\frac{2{P_{11}}{d_{3;23}}d_{2;12}^{2}\beta_{1}^{3}r_{L}^{{\d2}}{{\ln
}^2}{L_0}}{{L^2}\beta_{0}^{8}L_{0}^{3}}}
\cr &\hskip 2mm
- {\frac{2{P_{11}} {d_{5;13}}\beta_{1}^{3}r_{L}^{{\d3}}{{\ln
}^3}{L}}{{L^5}\beta_{0}^{8}}}
   \left( {d_{2;23}} + {d_{3;23}} \right)
+ {\frac{2{P_{11}}{d_{3;23}}{d_{6;13}} \beta_{1}^{4}r_{L}^{{\d3}}{{\ln
}^3}{L}}{{L^6}\beta_{0}^{10}}}
   \left( {d_{3;23}} + 2{d_{6;13}} \right)
-
{\frac{4{P_{11}}d_{3;12}^{3}d_{3;23}^{3}\beta_{1}^{4}r_{L}^{{\d2}}}{{L^3}\beta_{0}^{10}L_{0}^{3}}}
\cr &\hskip 2mm
- {\frac{2{P_{11}} \beta_{1}^{3}r_{L}^{\d1}{{\ln
}^3}{L_0}}{\beta_{0}^{8}L_{0}^{5}}}
   \left( {d_{2;23}}{d_{3;12}} + {d_{2;12}}{d_{3;23}} - {d_{2;23}}{d_{5;13}} -
{d_{3;23}}{d_{5;13}} \right)
- {\frac{2{P_{11}}{d_{3;23}}d_{3;12}^{3}\beta_{1}^{4}r_{L}^{{\d2}}{{\ln
}^2}{L_0}}{{L^3}\beta_{0}^{10}L_{0}^{3}}}
\cr &\hskip 2mm
- {\frac{ \beta_{1}^{2}r_{L}^{\d1}{{\ln }^3}{L_0}}{\beta_{0}^{6}L_{0}^{4}}}
   \left( {P_{10}}{d_{1;23}}{d_{3;12}} + {P_{01}}{d_{1;12}}{d_{3;23}} -
{P_{10}}{d_{1;23}}{d_{4;13}} - {P_{01}}{d_{3;23}}{d_{4;13}} \right)
}$$

$$\hskip -20pt\eqalign{
&\hskip 2mm
+ {\frac{2{d_{4;13}}\ln {L} \beta_{1}^{2}{r}_{L}^{{\d3}}}{{L^4}\beta_{0}^{6}}}
   \left( {P}_{11}^{(a)}{{{d}}_{1;23}}{d_{4;13}} +
4{P_{11}}{d_{2;23}}{d_{4;13}} + {{{P}}_{11}}{d_{3;23}}{d_{4;13}} +
2{P_{11}}d_{2;23}^{2} + {{{P}}_{11}}d_{3;23}^{2} -
2{P_{01}}{d_{4;13}}d_{3;23}^{2}
  \split
- {{{P}}_{01}}d_{3;23}^{3} - 3{P_{10}}{d_{1;23}}d_{4;13}^{2} -
3{P_{01}}{d_{3;23}}d_{4;13}^{2} \right)
\cr &\hskip 2mm
+ {\frac{2{d_{4;13}} \beta_{1}^{2}r_{L}^{{\d3}}}{{L^4}\beta_{0}^{6}}}
   \left( 2{P_{11}}{d_{4;13}}d_{2;23}^{2} + {P_{11}}{d_{4;13}}d_{3;23}^{2} +
{P_{11}}d_{3;23}^{3} - {P_{01}}{d_{4;13}}d_{3;23}^{3} +
P_{11}^{(a)}{d_{1;23}}d_{4;13}^{2} + 4{P_{11}}{d_{2;23}}d_{4;13}^{2}
   \split
  + {P_{11}}{d_{3;23}}d_{4;13}^{2} - 2{P_{01}}d_{3;23}^{2}d_{4;13}^{2} -
3{P_{10}}{d_{1;23}}d_{4;13}^{3} - 3{P_{01}}{d_{3;23}}d_{4;13}^{3} \right)
\cr &\hskip 2mm
- {\frac{4{P_{11}}{d_{5;13}}\ln {L}
\beta_{1}^{3}{r}_{L}^{{\d3}}}{{L^5}\beta_{0}^{8}}}
   \left( {{{d}}_{5;13}}d_{2;23}^{2} + 2{d_{5;13}}d_{3;23}^{2} + {d}_{3;23}^{3}
+ 3{d_{2;23}}d_{5;13}^{2} + 3{d_{3;23}}d_{5;13}^{2} \right)
\cr &\hskip 2mm
- {\frac{4{P_{11}}d_{5;13}^{2} \beta_{1}^{3}r_{L}^{{\d3}}}{{L^5}\beta_{0}^{8}}}
   \left( {d_{5;13}}d_{2;23}^{2} + 2{d_{5;13}}d_{3;23}^{2} + d_{3;23}^{3} +
3{d_{2;23}}d_{5;13}^{2} + 3{d_{3;23}}d_{5;13}^{2} \right)
\cr &\hskip 2mm
+ {\frac{4{P_{11}}{d_{3;23}}\ln {L}{d}_{6;13}^{2}
\beta_{1}^{4}{r}_{L}^{{\d3}}}{{L^6}\beta_{0}^{10}}}
   \left( 3{d_{3;23}}{d_{6;13}} + {d}_{3;23}^{2} + 6d_{6;13}^{2} \right)
+ {\frac{2{{\beta}_1} {d_{3;13}}r_{L}^{{\d3}}{{\ln
}^2}{L}}{{L^3}\beta_{0}^{4}}}
   \left( {P_{10}}{d_{1;23}} + {P_{01}}{d_{2;23}} \right)
\cr &\hskip 2mm
+ {\frac{{P_{00}}{d_{1;23}}{d_{2;13}}r_{L}^{{\d3}}{{\ln
}^2}{L}}{{L^2}\beta_{0}^{2}}}
+ {\frac{4{P_{11}}{d_{3;23}}
d_{6;13}^{3}\beta_{1}^{4}r_{L}^{{\d3}}}{{L^6}\beta_{0}^{10}}}
   \left( 3{d_{3;23}}{d_{6;13}} + d_{3;23}^{2} + 6d_{6;13}^{2} \right)
\cr &\hskip 2mm
+ {\frac{{d_{4;13}} \beta_{1}^{2}r_{L}^{{\d3}}{{\ln
}^2}{L}}{{L^4}\beta_{0}^{6}}}
   \left( P_{11}^{(a)}{d_{1;23}} + 4{P_{11}}{d_{2;23}} + {P_{11}}{d_{3;23}} -
3{P_{10}}{d_{1;23}}{d_{4;13}} - 3{P_{01}}{d_{3;23}}{d_{4;13}} -
2{P_{01}}d_{3;23}^{2} \right)
\cr &\hskip 2mm
- {\frac{2{P_{11}}{d_{5;13}} \beta_{1}^{3}r_{L}^{{\d3}}{{\ln
}^2}{L}}{{L^5}\beta_{0}^{8}}}
   \left( 3{d_{2;23}}{d_{5;13}} + 3{d_{3;23}}{d_{5;13}} + d_{2;23}^{2} +
2d_{3;23}^{2} \right)
\cr &\hskip 2mm
+ {\frac{2{P_{11}}{d_{3;23}}{d_{6;13}} \beta_{1}^{4}r_{L}^{{\d3}}{{\ln
}^2}{L}}{{L^6}\beta_{0}^{10}}}
   \left( 3{d_{3;23}}{d_{6;13}} + d_{3;23}^{2} + 6d_{6;13}^{2} \right)
+ {\frac{{P_{00}}{d_{1;23}} r_{L}^{\d1}{{\ln
}^2}{L_0}}{\beta_{0}^{2}L_{0}^{2}}}
   \left( {d_{1;12}} - {d_{2;13}} \right)
\cr &\hskip 2mm
+ {\frac{2{{\beta}_1} r_{L}^{\d1}{{\ln }^2}{L_0}}{\beta_{0}^{4}L_{0}^{3}}}
   \left( {P_{10}}{d_{1;23}}{d_{2;12}} + {P_{01}}{d_{1;12}}{d_{2;23}} -
{P_{10}}{d_{1;23}}{d_{3;13}} - {P_{01}}{d_{2;23}}{d_{3;13}} \right)
\cr &\hskip 2mm
+ {\frac{ \beta_{1}^{2}r_{L}^{\d1}{{\ln }^2}{L_0}}{\beta_{0}^{6}L_{0}^{4}}}
   \left( 4{P_{11}}{d_{2;12}}{d_{2;23}} + P_{11}^{(a)}{d_{1;23}}{d_{3;12}} +
P_{11}^{(b)}{d_{1;12}}{d_{3;23}} - P_{11}^{(a)}{d_{1;23}}{d_{4;13}} -
4{P_{11}}{d_{2;23}}{d_{4;13}}
  \split \hskip -20mm
- {P_{11}}{d_{3;23}}{d_{4;13}} - 2{P_{10}}{d_{1;23}}d_{3;12}^{2} -
2{P_{01}}{d_{1;12}}d_{3;23}^{2} + 2{P_{01}}{d_{4;13}}d_{3;23}^{2} +
3{P_{10}}{d_{1;23}}d_{4;13}^{2} + 3{P_{01}}{d_{3;23}}d_{4;13}^{2} \right)
\cr &\hskip 2mm
- {\frac{2{P_{11}} \beta_{1}^{3}r_{L}^{\d1}{{\ln
}^2}{L_0}}{\beta_{0}^{8}L_{0}^{5}}}
   \left( {d_{3;23}}d_{2;12}^{2} + {d_{3;12}}d_{2;23}^{2} -
{d_{5;13}}d_{2;23}^{2} + 2{d_{2;23}}d_{3;12}^{2} + 2{d_{2;12}}d_{3;23}^{2} -
2{d_{5;13}}d_{3;23}^{2}
  \split - 3{d_{2;23}}d_{5;13}^{2} - 3{d_{3;23}}d_{5;13}^{2} \right)
+ {\frac{{P_{11}}{d_{3;23}} \beta_{1}^{4}r_{L}^{\d1}{{\ln
}^4}{L_0}}{\beta_{0}^{10}L_{0}^{6}}}
   \left( {d_{3;12}} - {d_{6;13}} \right)
\cr &\hskip 2mm
+ {\frac{2{P_{11}}{d_{3;23}} \beta_{1}^{4}r_{L}^{\d1}{{\ln
}^2}{L_0}}{\beta_{0}^{10}L_{0}^{6}}}
   \left( 2{d_{3;23}}d_{3;12}^{2} + d_{3;12}^{3} + {d_{3;12}}d_{3;23}^{2} -
{d_{6;13}}d_{3;23}^{2} - 3{d_{3;23}}d_{6;13}^{2} - 6d_{6;13}^{3} \right)
\cr &\hskip 2mm
+ {\frac{2{P_{11}}{d_{3;23}} \beta_{1}^{4}r_{L}^{\d1}{{\ln
}^3}{L_0}}{\beta_{0}^{10}L_{0}^{6}}}
   \left( {d_{3;12}}{d_{3;23}} - {d_{3;23}}{d_{6;13}} + d_{3;12}^{2} -
2d_{6;13}^{2} \right)\,.
}\anoneqn$$

\listrefs
\bye